\documentclass[aps,prd,amsmath,amssymb]{revtex4}
\usepackage{bm}
\usepackage{indentfirst}
\usepackage{amsmath}
\usepackage{epsfig}
\usepackage{color}
\usepackage{amssymb}
\usepackage{amsfonts}
\usepackage{graphicx}
\usepackage{keyval,graphicx}
\usepackage{textcomp,wasysym}
\usepackage{subfigure}

\newcommand{\mev}{\mathrm{MeV}}
\newcommand{\gev}{\mathrm{GeV}}

\begin{document}

\title{Revisiting the pseudoscalar meson and glueball mixing and key issues in the search for pseudscalar glueball state}

\author{Wen Qin$^{1,3}$\footnote{{\it Email address:} qinwen@hunnu.edu.cn}, Qiang Zhao$^{2,3,4}$\footnote{{\it Email address:} zhaoq@ihep.ac.cn}, and Xian-Hui Zhong$^{1,3}$\footnote{{\it Email address:} zhongxh@hunnu.edu.cn}}

\affiliation{ 1) Department of Physics, Hunan Normal University, and Key Laboratory of
Low-Dimensional Quantum Structures and Quantum Control of Ministry
of Education, Changsha 410081, China }

\affiliation{ 2) Institute of High Energy Physics and Theoretical Physics Center for Science Facilities,
Chinese Academy of Sciences, Beijing 100049, China}

\affiliation{ 3) Synergetic Innovation Center for Quantum Effects and Applications (SICQEA),
Hunan Normal University, Changsha 410081, China}
\affiliation{ 4)  School of Physical Sciences, University of Chinese Academy of Sciences, Beijing 100049, China}

\date{\today}

\begin{abstract}

We revisit the mixing mechanism for pesudscalar mesons and glueball which is introduced by the axial vector anomaly. We demonstrate that the physical mass of the pseudoscalar glueball does not favor to be lower than 1.8 GeV if all the parameters are reasonably constrained. This conclusion, on the one hand, can accommodate the pseudoscalar glueball mass calculated by Lattice QCD, and on the other hand, is consistent with the high-statistics analyses at BESIII that all the available measurements do not support the presence of two closely overlapping pseudoscalar states in any exclusive channel. Such a result is in agreement with the recent claim that the slightly shifted peak positions for two possible states $\eta(1405)$ and $\eta(1475)$ observed in different channels are actually originated from one single state with the triangle singularity interferences. By resolving this long-standing paradox, one should pay more attention to higher mass region for the purpose of searching for the pseudoscalar glueball candidate.

\end{abstract}


\pacs{}
\maketitle

\section{Introduction}

The non-Abelian property of Quantum Chromo-Dynamics (QCD) predicts the possible existence of glueball states as a peculiar manifestation of the strong interaction in the non-perturbative regime. However, until now indisputable experimental evidence for the glueball states is still lacking. In the pseudoscalar sector, the flux tube model supports a low-lying pseudoscalar glueball with a mass around 1.4 GeV~\cite{Faddeev:2003aw}. This was the mass region accessible by several experiments in 1980's and 1990's, for instance, Mark-III~\cite{Bai:1990hs,Bolton:1992kb}, DM-2~\cite{Augustin:1989zf,Augustin:1990ki}, OBELIX~\cite{Bertin:1995fx,Bertin:1997zu,Cicalo:1999sn}, and BES-II~\cite{Bai:2004qj}. Reviews on the early experimental observations can be found in Refs.~\cite{Masoni:2006rz,Klempt:2007cp}. With the strong motivation of looking for glueball candidates in experiment, the observation of three possible pseudoscalar states with isospin 0 around 1.3$\sim$ 1.5 GeV, i.e. $\eta(1295)$, $\eta(1405)$, $\eta(1475)$, was regarded as the clues for the presence of a pseudoscalar glueball in association with the isospin singlets in the $q\bar{q}$ scenario. Note that there have been well-established states, i.e. $\pi(1300)$ and $K(1460)$, in the same mass region with which the first radial excitation of the $q\bar{q}$ pseudoscalar meson nonet with $J^P=0^-$ can be formed~\cite{Klempt:2007cp,Olive:2016xmw,Yu:2011ta}. For a long time following the rather vague experimental results, there have been tremendous efforts trying to understand the property of these three states among which the $\eta(1405)$ has been assigned as the most-likely pseudoscalar glueball candidate. Other explanations for the out-numbering of isoscalar pseudoscalar states around $1.3\sim 1.5$ GeV include dynamically generated states~\cite{Albaladejo:2010tj} and tetraquarks~\cite{Fariborz:2009cq}. However, any explanation for the out-numbering problem should first confirm whether indeed an additional state is present.

The phenomenological studies of the pseudoscalar glueball candidate $\eta(1405)$ have been focussed on the following main issues:
\begin{itemize}
\item
Whether there are mixings among the ground state pseudoscalar mesons $\eta$ and $\eta'$, and the pseudoscalar glueball? And how to disentangle their internal structures? What are the consequences from such state mixings~\cite{Cheng:2008ss,Close:1996yc,Li:2007ky,Gutsche:2009jh,Li:2009rk,Eshraim:2012jv}?

\item
What causes the low mass of pseudoscalar glueball compared with the lattice QCD (LQCD) calculations~\cite{Morningstar:1999rf,Bali:1993fb,Chen:2005mg,Chowdhury:2014mra,Richards:2010ck,Sun:2017ipk}?

\item
What is the role played by the triangle singularity mechanism arising from the rescattering of $K\bar{K}^*+c.c.$ to different final states in $\eta(1405/1475)\to K\bar{K}^*+c.c.\to K\bar{K}\pi$, $\eta\pi\pi$ and $3\pi$~\cite{Zhao:2017wey,Wu:2011yx,Wu:2012pg,Aceti:2012dj}?

\end{itemize}

Following the questions from item one, most studies assume certain mixing mechanisms among $\eta$, $\eta'$ and $\eta(1405)$ and investigate the properties of $\eta(1405)$ in gluon-rich processes such as $J/\psi$ radiative decays. Also, the gluon contents inside $\eta$ and $\eta'$ can provide some hints of glueball states due to the mixing mechanism. As a consequence of such a mixing, one expects that observable effects can be measured in experiment which can make the glueball state different from the $q\bar{q}$ mesons. However, it is still difficult to conclude that the pseudoscalar glueball state has been observed in experiment taking into account the high precision measurements from the BESIII experiment and LQCD calculations. This is related to the questions raised above in items two and three.

During the past decade the progress of LQCD has brought many novel insights into the light hadron spectroscopy via numerical simulations of the non-perturbative strong interactions. Interesting and surprisingly, it shows that the lightest pseudoscalar glueball should have a mass around 2.4$\sim$2.6 GeV in a quenched calculation~\cite{Chowdhury:2014mra,Chen:2005mg,Bali:1993fb,Morningstar:1999rf}, while the later dynamical calculations~\cite{Richards:2010ck,Sun:2017ipk} suggest that the mass of the lightest pseudoscalar glueball does not change much compared with the quenched result. This is obviously in contradiction with the data if $\eta(1405)$ is assigned as a glueball candidate.

In parallel with the LQCD studies, great efforts have been made in experiment in order to establish $\eta(1405)$ as an additional state apart from $\eta(1295)$ and $\eta(1475)$. A natural expectation is that since both $\eta(1405)$ and $\eta(1475)$ have the same quantum number and can couple to the same hadronic final states there should be channels that they both can have observable couplings, hence, nontrivial structures caused by two closely overlapping and interfering states should appear in the mass spectra. However, with the high-statistics measurements in various channels, e.g. in $J/\psi$ radiative and hadronic decays at BESIII{~\cite{BESIII:2012aa,Ablikim:2011pu,Ablikim:2010au}}, there is no any evidence indicating that two nearby $\eta(1405)$ and $\eta(1475)$ have been produced together in the same channel. All the data so far only show one peak structure around 1.42 GeV and no need to introduce interfering states from two nearby states. These new measurements actually have brought serious questions on the need for an additional $\eta(1405)$ apart from the radial excitation of the $q\bar{q}$ isoscalars $\eta(1295)$ and $\eta(1475)$.

The new data also raise new features for the radial excitation spectrum of the isoscalar states $\eta$ and $\eta'$. One notices that the single peak positions for $\eta(1405/1475)$ are slightly shifted in different channels. In particular, the observation of the significantly large isospin breaking effects in $J/\psi\to\gamma \eta(1405/1475)\to\gamma +3\pi$ can be regarded as an indication of a special mechanism that causes the mysterious phenomena around 1.4$\sim$1.5 GeV for the isoscalar pseudoscalar meson spectrum~\cite{BESIII:2012aa}. It was proposed by Refs.~\cite{Wu:2011yx,Wu:2012pg} that the presence of the so-called ``triangle singularity (TS)" mechanism can enhance the isospin breaking effects and shift the peak positions of a single state by the interferences in exclusive decay channels. Similar analysis of Ref.~\cite{Aceti:2012dj} also confirms that the TS contribution is needed in order to understand the strong isospin breaking effects.

The TS mechanism was first investigated by Landau in 1950's~\cite{Landau:1959fi} and followed up by many detailed studies later\cite{Cutkosky:1960sp,bonnevay:1961aa,Peierls:1961zz,Goebel:1964zz,hwa:1963aa,landshoff:1962aa,Aitchison:1969tq}. It states that for an initial state with energies near an intermediate open threshold, if the rescattering between these two intermediate states by exchanging another state (i.e. via a triangle diagram) into three-body final states would allow such kinematics that all the three intermediate states can approach their on-shell condition simultaneously, to be located within the physical region, then the triangle loop amplitude will be enhanced by the three-body singularity as the leading contribution. As a consequence, its interference with the tree-level transition amplitude of the initial state can shift the its peak position and even change the lineshape~\cite{Wu:2011yx}. In the case of $\eta(1405/1475)\to 3\pi$ the mass of the initial state $\eta(1405/1475)$ is within the TS kinematic region and has strong couplings to $K\bar{K}^*+c.c.$  Thus, the intermediate $K\bar{K}^*+c.c.$ and the exchanged kaon in the triangle loop can approach the on-shell condition simultaneously and results in the strong enhancement of the isospin breaking on top of the $a_0(980)$ and $f_0(980)$ mixing. The recognition of the TS mechanism here provides an alternative explanation for understanding the $\eta(1405)$-$\eta(1475)$ puzzle and can resolve the contradiction between the LQCD results and experimental observations for the pseudoscalar glueball. Recent detailed analyses and discussions on the TS mechanism can be found in Refs.~\cite{Liu:2015taa}. More recognitions of this special kinematic effects in various processes can be found in the literature~\cite{Wang:2013cya,Wang:2013hga,Liu:2013vfa,Liu:2014spa,Szczepaniak:2015eza,Guo:2015umn,Liu:2015fea,Zhao:2016akg,Bayar:2016ftu, Wang:2016dtb,Liu:2017vsf,Xie:2017mbe,Sakai:2017iqs,Pavao:2017kcr,Sakai:2017hpg} and recent reviews~\cite{Zhao:2017wey,Guo:2017jvc}.

The above progress suggests that the pseudoscalar meson and glueball mixing mechanism should be re-investigated. Moreover, given that the pseudoscalar glueball mass in the quenched approximation is around 2.4$\sim$2.6 GeV, its mixing with the $c\bar{c}(0^{-+})$ should also be considered. An earlier study of the mixing mechanism has implemented the anomalous Ward identities with the corresponding equations of motion which connect the transition matrix elements of vacuum to $\eta$, $\eta'$ and glueball  to the pseudoscalar densities and the U(1) anomaly~\cite{Tsai:2011dp}. There, the physical glueball state was assigned to $\eta(1405)$ and then the mixing effects on $\eta$, $\eta'$ and $\eta_c$ were studied. Due to a large number of parameters in the mixing scheme of Ref.~\cite{Tsai:2011dp}, it shows that a re-investigation of the parameter space is necessary. In particular, a detailed analysis of the sensitivity of the glueball mass range to the mixing parameters is necessary. This will help further clarify the puzzling situation around 1.4 GeV in the $I=0$ pseudoscalar spectrum. One also notices that the recent analysis of Ref.~\cite{Mathieu:2009sg} in a chiral Lagrangian approach with an axial anomaly coupling also leads to a much heavier glueball mass than the range of around 1.4 GeV.

As follows, we first introduce the formulation of the mixing scheme via the axial vector anomaly as studied in Refs.~\cite{Feldmann:1998vh,Feldmann:1998sh} in  Sec.~\ref{model}. We then inspect the parameter space and impose constraints on these parameters in order to investigate the mass range of the pseudoscalar glueball.  In particular, the sensitivities of the physical glueball mass to the parameters will be scrutinized.  In Sec.~\ref{result-discuss}, the numerical results are presented and discussed. A brief summary is given in Sec.~\ref{summary}.

This work is organized as follows: In Sec.~\ref{model} we first introduce the formulation of the mixing scheme via the axial vector anomaly as studied in Refs.~\cite{Feldmann:1998vh,Feldmann:1998sh}. We inspect the parameter space and try to investigate the sensitivities of the physical glueball mass to the parameters.  In Sec.~\ref{result-discuss}, the numerical results are presented and discussed. A brief summary is given in Sec.~\ref{summary}.

\section{The mixing formalism}\label{model}

\subsection{$\eta-\eta^{\prime}-G-\eta_c$ mixing scheme}

As stated in Ref.~\cite{Feldmann:1998vh,Feldmann:1998sh}, the well-known axial vector anomaly is,
\begin{equation}
\label{anomaly }
\partial^{\mu} J^j_{\mu 5} =\partial^\mu (\bar j \gamma_\mu \gamma_5 j) = 2 m_j (\bar j i \gamma_5 j) + \frac{\alpha_s}{4 \pi} G \tilde{G}
\end{equation}
where $j$ denotes the $q,s,c$ quark respectively, and $m_j$ denotes the {quark masses}, $G$ and $\tilde G$ denote the strength tensor and the dual of the gluon field.
 The physical states are mixture of the pure states via a unitary matrix $U$ as,
 \begin{equation}
\label{ }
\begin{pmatrix}
     | \eta \rangle    \\
    | \eta^\prime \rangle  \\
    | G  \rangle   \\
    | \eta_c \rangle
\end{pmatrix}  = U
\begin{pmatrix}
     | \eta_q \rangle    \\
    | \eta_s \rangle  \\
    | g  \rangle   \\
    | \eta_Q \rangle
\end{pmatrix}
\end{equation}
where $| \eta_q \rangle    ,
    | \eta_s \rangle ,
    | g  \rangle   ,
    | \eta_Q \rangle$ denote $| qq \rangle \equiv | ( u \bar u + d \bar d) / \sqrt{2}  \rangle$, $|s \bar s\rangle$,  the unmixed glueball state, and the unmixed heavy quark state $|c \bar c\rangle$.

Assuming that the decay constants in the flavor basis follow the same mixing pattern of the particle states~\cite{Feldmann:1998vh}, we have
\begin{equation}
\label{ }
\setlength{\arraycolsep}{5pt}
\begin{pmatrix}
    f_\eta^q  &   f_\eta^s   &   f_\eta^c   \\
     f_{\eta^\prime}^q  &   f_{\eta^\prime}^s   &   f_{\eta^\prime}^c   \\
    f_G^q  &   f_G^s   &   f_G^c   \\
         f_{\eta_c}^q  &   f_{\eta_c}^s   &   f_{\eta_c}^c \end{pmatrix}  = U
\begin{pmatrix}
    f_q  &  0  &   0  \\
     0    &  f_s  &   0  \\
       0    &  0  &   0  \\
    0    &  0  &   f_c
\end{pmatrix}
\end{equation}
where all the OZI-suppressed off-diagonal elements are neglected.

The pseudscalar meson decay constants are defined as follows,
\begin{equation}
\label{ }
\langle 0| \partial^\mu J_{\mu5}^j |P\rangle= M_P^2 f_P ^j \ ,
\end{equation}
where $M_P$ is the diagonal mass matrix of the physical states that is explicitly written as,
\begin{equation}
\begin{pmatrix}
    M_\eta^2  &  0  &   0 &  0 \\
     0    &  M_{\eta^\prime}^2  &   0  &  0  \\
       0    &  0 &  M_G^2   &   0  \\
    0    &  0  & 0   &   M_{\eta_c}^2
\end{pmatrix} \ .
\end{equation}
Noted that the meson state is with the dimension of mass$^{-1}$ and the decay constant with the dimension of mass.

Based on the above definitions and assumption, we can obtain the mass matrix on the flavor basis from two ways. On the one hand, the mass matrix on the flavor basis is related to the physical particle mass via a unitary transformation. On the other hand, according to the definition of the decay constants, the mass matrix is also related to the axial vector current divergences in a more dynamical and explicit way. Although some of the matrix elements cannot be well constrained and determined quantitatively, they are not going to affect our discussions here due to their small values that can be qualitatively determined. The mass matrix in terms of the physical masses can be written as,
\begin{equation}
\label{M-angle}
\mathcal{M}_{qsgc}=U^\dag M_P^2 U \ .
\end{equation}

In order to obtain the mass matrix in terms of the divergences of the axial vector current, we firstly define the following abbreviations for pseudoscalar densities and the U(1) anomaly matrix elements as done in Ref.~\cite{Tsai:2011dp}:
\begin{eqnarray}
m_{qq,qs,qg,qc}^2&\equiv&\frac{\sqrt{2}}{f_q}\langle 0|m_u\bar u
i\gamma_5 u+m_d\bar d
i\gamma_5 d|\eta_q,\eta_s,g,\eta_Q\rangle\;,\nonumber\\
m_{sq,ss,sg,sc}^2&\equiv&\frac{2}{f_s}\langle 0|m_s\bar s i\gamma_5
s|\eta_q,\eta_s,g,\eta_Q\rangle, \nonumber\\
m_{cq,cs,cg,cc}^2&\equiv&\frac{2}{f_c}\langle 0|m_c\bar c i\gamma_5
c|\eta_q,\eta_s,g,\eta_Q\rangle, \nonumber\\
G_{q,s,g,c}&\equiv &\frac{\alpha_s}{4\pi}\langle 0|G{\tilde
G}|\eta_q,\eta_s,g,\eta_Q\rangle. \label{mqq}
\end{eqnarray}
Note that the definition for $q \ (u, \ d)$ quark current is different from other quark flavors by a factor of $\sqrt{2}$ due to the definition of the $| q\bar q\rangle$.

Then, the mass density matrix of the $q,s,c$ dimension can be written explicitly in a dynamical way as in Ref.~\cite{Tsai:2011dp},
\begin{equation}
\label{Mdynexp}
\setlength{\arraycolsep}{8pt}
\mathcal{\tilde{M}}_{qsgc}=\begin{pmatrix}
m_{qq}^2+\sqrt{2}G_q/f_q & m_{sq}^2+G_q/f_s & m_{cq}^2+G_q/f_c\\
m_{qs}^2+\sqrt{2}G_s/f_q & m_{ss}^2+G_s/f_s & m_{cs}^2+G_s/f_c\\
m_{qg}^2+\sqrt{2}G_g/f_q & m_{sg}^2+G_g/f_s & m_{cg}^2+G_g/f_c\\
m_{qc}^2+\sqrt{2}G_c/f_q & m_{sc}^2+G_c/f_s & m_{cc}^2+G_c/f_c
\end{pmatrix} \ .
\end{equation}
The mass density matrix obtained from these two ways should be the same. Thus, the mixing information could be revealed.

To proceed, we first analyze the parameters involved in this mixing scheme by looking at the transformation matrix $U$. In Ref.~\cite{Peng:2011ue}, a general form for the unitary mixing matrix is presented with six independent rotation angles. It would not be realistic to determine all of them based on what we know about the pseudoscalar meson and glueball mixing. In order to implement constraints on the mixing matrix elements, we take a similar strategy of Ref.~\cite{Tsai:2011dp} to reduce the number of parameters.

Firstly, the mixing between the light favor octet state $\eta_8$ and glueball is neglected in the SU(3) flavor symmetry.  Secondly, the heavy-flavor state mixing with the light-flavor state is also neglected, since they have a large mass difference and is OZI suppressed. These will reduce the number of undetermined parameters to only three mixing angles, i.e. the mixing angle $\phi_Q$ between the heavy-flavor state and glueball, the mixing angle $\phi_G$ for the glueball and light-flavor singlet state mixing, and the mixing angle $\theta$ between the octet and singlet light flavor states that mainly determines the structure of $\eta$ and $\eta^\prime$.  So the mixing matrix between the flavor states and the physical states can be written as~\cite{Tsai:2011dp},
\begin{eqnarray}
\label{Umatrix}
U(\theta,\phi_G,\phi_Q)&=&U_{34}(\theta)U_{14}(\phi_G)U_{12}(\phi_Q)U_{34}(\theta_i),\nonumber\\
&=&
\setlength{\arraycolsep}{5pt}
\begin{pmatrix}
      c\theta c\theta_i-s\theta c\phi_G s\theta_i & -c\theta s\theta_i-s\theta c\phi_G c\theta_i
& -s\theta s\phi_G c\phi_Q& -s\theta s\phi_G s\phi_Q\\
s\theta c\theta_i+c\theta c\phi_G s\theta_i & -s\theta
s\theta_i+c\theta c\phi_G c\theta_i & c\theta s\phi_G c\phi_Q& c\theta s\phi_G s\phi_Q\\
-s\phi_G s\theta_i &-s\phi_G c\theta_i & c\phi_G
c\phi_Q & c\phi_G s\phi_Q \\
0 &0 & -s\phi_Q & c\phi_Q
\end{pmatrix} \ ,
\end{eqnarray}
where $c$ and $s$ are the short-handed notations for ``$\cos$" and ``$\sin$"; $\theta_i$ are the ideal mixing angle between $\eta_q\equiv (u\bar{u}+d\bar{d})/\sqrt{2}$ and $\eta_s\equiv s\bar{s}$. The mass density matrix element from the physical state mass through the $U$ matrix can be obtained. The explicit expressions for each matrix element can be found in Ref.~\cite{Tsai:2011dp}. Here, we concentrate on the matrix elements that are relevant in the extraction of physical quantities of interest.

\subsection{Constrain the parameters}

The mixing mechanism discussed in Ref.~\cite{Tsai:2011dp} and summarized above allows us to express the mass matrix as follows:
\begin{equation}
\label{Matrix-EQ}
U^\dag \begin{pmatrix}
    M_\eta^2  &  0  &   0 &  0 \\
     0    &  M_{\eta^\prime}^2  &   0  &  0  \\
       0    &  0 &  M_G^2   &   0  \\
    0    &  0  & 0   &   M_{\eta_c}^2
\end{pmatrix} U
\begin{pmatrix}
   1  &  0  &   0 \\
     0    &  1  &   0    \\
       0    &  0 &  0     \\
    0    &  0  & 1
\end{pmatrix}
=\setlength{\arraycolsep}{8pt}
\begin{pmatrix}
m_{qq}^2+\sqrt{2}G_q/f_q & m_{sq}^2+G_q/f_s & m_{cq}^2+G_q/f_c\\
m_{qs}^2+\sqrt{2}G_s/f_q & m_{ss}^2+G_s/f_s & m_{cs}^2+G_s/f_c\\
m_{qg}^2+\sqrt{2}G_g/f_q & m_{sg}^2+G_g/f_s & m_{cg}^2+G_g/f_c\\
m_{qc}^2+\sqrt{2}G_c/f_q & m_{sc}^2+G_c/f_s & m_{cc}^2+G_c/f_c
\end{pmatrix} \ .
\end{equation}
On the left hand side of the equation, there are four parameters, i.e. the physical glueball mass $M_G$ and three mixing angles. On the right hand side, more parameters emerge which are related to the mixing dynamics. Apart from $f_q$, $f_s$, $f_c$, $m_{cc}$, $m_{qq}$, $m_{ss}$ that are more explicit and can be estimated phenomenologically by observable physical quantities,  there are still nine pseudoscalar densities and four U(1) anomaly matrix elements to be determined. Note that parameters $m_{qq}$ and $m_{ss}$ are related to the relatively well defined $\eta$ and $\eta^\prime$ mixing, they can be extracted from $\mathcal{\tilde{M}}_{qsgc}^{11}$ and $\mathcal{\tilde{M}}_{qsgc}^{22}$ in Ref.~\cite{Tsai:2011dp}. Besides, $m_{qq}$ is too small and in some cases the result even flips the sign~\cite{Cheng:2008ss}. Actually, since the masses of $\eta$ and $\eta'$ are rather far away from the glueball mass, the mixing effects due to the presence of glueball are expected to be small. In this sense, the constraint from the glueball contents of $\eta$ and $\eta'$ could be still marginal.

As discussed in Ref.~\cite{Cheng:2008ss},  the OZI-violating light-flavor pseudoscalar density $m_{qs}, m_{sq}$ scales as $O(1/N_c)$ in the limit of the large color number $N_c$, and $m_{qg}$ is of the order higher than $m_{qq}$ which is as small as $m_{\pi}^2$. Thus, we can drop these three parameters in this analysis and this is different from the treatment of Ref.~\cite{Tsai:2011dp}. We will show later that this is a reasonable assumption.

The flavor mixing angle $\theta$ for $\eta$ and $\eta^\prime$ are constrained in a finite range $-17^\circ < \theta <-11^\circ$. Although $\theta$ is not precisely fixed, its influence on the glueball property is rather small. A relatively small mixing angle $\theta<-10^\circ$ is favored by the unquenched LQCD calculation~\cite{Fukaya:2015ara}. Thus, we adopt $\theta=-11^\circ$ which is the same as in Ref.~\cite{Tsai:2011dp}.

The glueball component within the physical $\eta_c$ can be estimated by an empirical gluon power counting rule~\cite{Close:2005vf} to combine with the experimental data of the branching ratio of $\eta_c \to \gamma \gamma$, as done in Ref.~\cite{Tsai:2011dp}. With the updated experimental data $BR(\eta_c \to \gamma \gamma)=(1.59\pm0.13)\times 10^{-4}$~\cite{Olive:2016xmw}, $\phi_Q=-2.7^\circ$ and $11.6^\circ$ are obtained with the central value. As discussed in Ref.~\cite{Tsai:2011dp}, a negative $\phi_Q$ is not favored by the radiative decays of $J/\psi, \ \psi^\prime \to \gamma \eta_c$. Therefore, we adopt the positive value of $\phi_Q=11.6^\circ$.

The last mixing angle $\phi_G$ is directly related to the physical glueball mass as shown by the third row of the mass density matrix (Eq.~(\ref{Matrix-EQ})). Thus, a reliable determination of this quantity is crucial for estimating the physical glueball mass range.

The above consideration has significantly reduced the parameter number but still there are more than 10 parameters to be determined in Eq.~(\ref{Matrix-EQ}). In Refs.~\cite{Cheng:2008ss,Tsai:2011dp}, a different treatment for the parameters was applied to estimate the glueball mass. By taking the ratio of elements, e.g. $\mathcal{\tilde{M}}_{qsgc}^{31}/\mathcal{\tilde{M}}_{qsgc}^{32}$ in Eq.~(\ref{Matrix-EQ}), and assuming the negligibly small values of $m_{qg}^2$ and $m_{sg}^2$ compared with $\sqrt{2} G_g/f_q$ and $G_g/f_s$, the glueball mass will depend on the ratio of $f_s/f_q$, while its dependence on $G_g$ will be cancelled. A caveat of this treatment is that $m_{sg}^2$ actually is not small enough to be neglected. This point will be discussed later.  On the other hand, if $m_{qg}^2$ and $m_{sg}^2$ are neglected, it will lead to independence of the glueball mass on parameters $G_{q,s,g,c}$. However, since $G_{q,s,g,c}$ describes the contributions from the pseudoscalar U(1) anomaly in Eq.~(\ref{Matrix-EQ}), one would expect its direct connection with the physical glueball mass in the constraint relation.  Our revisit to this issue is to examine how the glueball mass should depend on $G_{q,s,g,c}$ in an explicit way.

Still focussing on $\mathcal{\tilde{M}}_{qsgc}^{31}/\mathcal{\tilde{M}}_{qsgc}^{32}$ in Eq.~(\ref{Matrix-EQ}), we extend the discussions on the parameters slightly. These two elements have the following expressions:
\begin{eqnarray}
\label{m31}
\mathcal{\tilde{M}}_{qsgc}^{31}&=&m_{qg}^2+
\sqrt2 G_g/f_q\nonumber\\
&=&-M_\eta^2(c\theta c\theta_i-s\theta c\phi_G s\theta_i)
s\theta s\phi_G c\phi_Q\nonumber\\
& &+M_{\eta'}^2(s\theta c\theta_i+c\theta c\phi_G s\theta_i)
c\theta s\phi_G c\phi_Q
-M_G^2 c\phi_Gs\phi_G s\theta_i c\phi_Q,
\end{eqnarray}
and
\begin{eqnarray}
\label{m32}
\mathcal{\tilde{M}}_{qsgc}^{32}&=& m_{sg}^2 + G_g/f_s\nonumber\\
             &=& M_\eta^2(c\theta s\theta_i+s\theta c\phi_G c\theta_i)
s\theta s\phi_G c\phi_Q\nonumber\\
& &+M_{\eta'}^2(-s\theta s\theta_i+c\theta c\phi_G c\theta_i)
c\theta s\phi_G c\phi_Q
-M_G^2 c\phi_Gs\phi_G c\theta_i c\phi_Q
\end{eqnarray}
Note that in Eq.~(\ref{m31}) it is safe to neglect $m_{qg}^2$ and only keep term $\sqrt2 G_g/f_q$ since $m_{qg}^2\ll m_{qq}^2$ with $m_{qq}^2$ about 36 times smaller than $\sqrt2 G_g/f_q$. However, it is not obvious to neglect $m_{sg}^2$ in Eq.~(\ref{m32}) since so far we only know the relation of $m_{sg}\ll m_{ss}$~\cite{Cheng:2008ss}, but have no information about the values of $m_{sg}$. Similar situation occurs with $m_{cg}^2$ and $m_{cc}^2$ when treating the elements $\mathcal{\tilde{M}}_{qsgc}^{41}$ and $\mathcal{\tilde{M}}_{qsgc}^{42}$, i.e. the only known information is $m_{cg}\ll m_{cc}$. Apparently, if the value of $m_{sg}^2$ is compatible with $G_g/f_s$, it will result in large uncertainties when taking the ratio of $\mathcal{\tilde{M}}_{qsgc}^{31}/\mathcal{\tilde{M}}_{qsgc}^{32}$. Meanwhile, the sensitivities of the glueball mass to $G_g$ will be lost. This problem can be seen more clearly if one compares the following two equal ratios extracted from Eq.~(\ref{Matrix-EQ}):
\begin{eqnarray}\label{R31-32}
{\hat R}_{31/32}&\equiv &\frac{\mathcal{\tilde{M}}_{qsgc}^{31}}{\mathcal{\tilde{M}}_{qsgc}^{32}}=\frac{m_{qg}^2+\sqrt{2}G_g/f_q}{m_{sg}^2+G_g/f_s} \ ,
\end{eqnarray}
and
\begin{eqnarray}\label{R41-42}
{\hat R}_{41/42}&\equiv &\frac{\mathcal{\tilde{M}}_{qsgc}^{41}}{\mathcal{\tilde{M}}_{qsgc}^{42}}=\frac{m_{qc}^2+\sqrt{2}G_c/f_q}{m_{sc}^2+G_c/f_s} \  ,
\end{eqnarray}
where Eq.~(\ref{R31-32}) leads to ${\hat R}_{31/32}\simeq \sqrt{2}f_s/f_q$ after neglecting $m_{qg}^2$ and $m_{sg}^2$. However, note that $\sqrt{2}|G_c/f_q|\simeq 0.039$ GeV$^2$ and $G_c/f_s\simeq 0.023$ GeV$^2$ both are much smaller than $|m_{qc}^2|=1.197$ GeV$^2$ and $|m_{sc}^2|=0.092$ GeV$^2$ in Eq.~(\ref{R41-42}). The neglect of $m_{qg}^2$ and $m_{sg}^2$ in Eq.~(\ref{R31-32}) and $m_{qc}^2$ and $m_{sc}^2$ in Eq.~(\ref{R41-42}) cannot be justified. Thus, although the equivalence ${\hat R}_{31/32}={\hat R}_{41/42}$ can be deduced rigorously from the left-hand side of Eq.~(\ref{Matrix-EQ}), the relation of ${\hat R}_{31/32}={\hat R}_{41/42}\simeq \sqrt{2}f_s/f_q$ actually does not hold.

To proceed, we take a slightly different strategy to determine the parameters and extract the pseudoscalar glueball mass. First, it should be noted that an explicit relation between $G_g$ and the glueball mass should be retained. Note that the value $|G_g|= (0.054 \pm 0.008) \ \gev^3$, has been calculated by lattice QCD in the quenched approximation~\cite{Chen:2005mg}. Although the sign of $G_g$  is not determined by LQCD, we will show that the positive value can be excluded since it will lead to negative values for the glueball mass. We also take the decay constant $f_q$ as an input. It is relatively well constrained to be $(1\sim1.1) f_\pi$ while $f_s$ varies within a range of $(1.3 \sim 1.6) f_\pi$~\cite{Feldmann:1999uf,Escribano:2007cd,Feldmann:1998vh,Feldmann:1998sh}. Another two parameters that we adopt are $f_c=487.4$ MeV \cite{Peng:2011ue} and $m_{cc}\approx M_{\eta_c}^2$~\cite{Feldmann:1998vh}. These are reasonable approximations taking into account the success of potential model in the description of low-lying charmonium states. Note that the decay constant $f_{J/\psi}=405 \ \mev$~\cite{Davies:2013ju} and the quenched mass $m_{\eta_c}=3.024\ \gev$~\cite{Dudek:2007wv} are provided by LQCD. As the leading approximation we assume that the $c\bar{c}$ bare vector ($J/\psi$) and bare pseudoscalar ($\eta_c$) share the same wave function at the origin as expected in the heavy quark spin symmetry (HQSS) limit, although in reality the HQSS breaking effects cannot be neglected. Following the same reason, it is reasonable to adopt the physical $\eta_c$ mass for $m_{cc}$ in contrast with the LQCD quenched mass $m_{\eta_c}=3.024\ \gev$.

With the above parameters fixed we are left with 12 equations with 13 undetermined parameters, i.e. $M_G, \ \phi_G, \ G_q, \ G_s, \ G_c, \ m_{sg}, \ m_{qc}, \ m_{sc}, \ m_{cq}, \ m_{cs}, \ m_{cg}, \ m_{qq}, \ m_{ss}$.  Note that as mentioned earlier, $m_{sq}^2,m_{qs}^2,m_{qg}^2$ are neglected since $m_{qs,sq}^2 \ll m_{qg}^2\ll m_{qq}^2$ with $m_{qq}^2$ about 36 times smaller than $\sqrt2 G_g/f_q$. Therefore, we make the approximation to Eq.~(\ref{m31}) which leads to
\begin{eqnarray}
\label{MG}
M_G^2&=&-\frac{1}{\cos\phi_G\sin\theta_i \cos\phi_Q}\{\frac{\sqrt2 G_g/f_q}{\sin\phi_G}-[-M_\eta^2(\cos\theta \cos\theta_i-\sin\theta \cos\phi_G \sin\theta_i)
\sin\theta  \cos\phi_Q\nonumber\\
& &+M_{\eta'}^2(\sin\theta \cos\theta_i + \cos\theta \cos\phi_G \sin\theta_i)
\cos\theta \cos\phi_Q]\} \ .
\end{eqnarray}
One notices that not all the parameters are explicitly correlated in a single relation in this mixing scheme. This allows us to investigate the relation between two unknown quantities in a single equation while the other parameters can be fixed with reasonable values. Following this consideration, Eq.~(\ref{MG}) can be approximated by
\begin{eqnarray}
\label{MG-1}
M_G^2 &\approx&-\frac{1}{\sin\theta_i }\{\frac{\sqrt2 G_g/f_q}{\sin\phi_G} - M_{\eta'}^2 \sin\theta_i - (M_{\eta'}^2-M_\eta^2)\sin\theta\cos(\theta+\theta_i)\}
\ ,
\end{eqnarray}
where all the cosine values of the small angles have been taken as unity, and the glueball mass sensitivity to $\phi_G$ can be investigated. Note that $\phi_G$ is not well constrained and its value varies in a wide range, depending on the parametrization of the mixing matrix, experimental inputs and fitting procedures~\cite{Cheng:2008ss}. For example, $(12\pm13)^\circ$ is obtained in  the radiative decay of $V \to P \gamma, P \to V\gamma$  independent of $f_q$ and $f_s$~\cite{Escribano:2007cd}. The variation range of $\phi_G$ was found to be $(32^{+11}_{-22})^\circ$ in the strong process $J/\psi \to V P$ in Ref.~\cite{Escribano:2008rq}. Similar scenario was also studied in Refs.~\cite{Zhao:2006gw,Li:2007ky}.

In the next Section we provide relations for the other parameters in terms of $\phi_G$ and the ratios ${f_c}/{f_q}$, ${ f_s}/{f_q}$, and ${f_s}/{f_c}$. Although these three ratios are not independent, the idea is to investigate whether those undetermined parameters can have acceptable values located within a common regime of $\phi_G$ and the ratios.

Note that for the determination of the pseudoscalar glueball mass via Eq.~(\ref{m31}), the decay constant $f_s$ is not explicitly involved. However, its correlation with other parameters will affect the result to some extent, in particular, via the ratio of $f_s/f_q$. Note that  $f_q\simeq (1\sim1.1) f_\pi$ is relatively well determined while $f_s\simeq (1.3 \sim 1.6) f_\pi$ is well estimated~\cite{Feldmann:1999uf,Escribano:2007cd,Feldmann:1998vh,Feldmann:1998sh}. In order to determine all the parameters self-consistently, we will fix the ratio of $f_s/f_q$ with commonly accepted values and solve the 12 equations with 12 parameters. The ratio of $f_s/f_q$ contains the uncertainties of SU(3) symmetry breaking effect. We will show later that the uncertainties arising from the SU(3) symmetry breaking will not change the magnitude hierarchy of the correlated parameters. In particular, we will see that the calculated glueball mass should not be sensitive to the ratio $f_s/f_q$ which is different from the result of Refs.~\cite{Cheng:2008ss,Tsai:2011dp} due to different ways of treating the parameters.

\section{Results and discussion}\label{result-discuss}

\subsection{Pseudoscalar glueball mass and its correlations with other parameters}

We first study the relation between $M_G$ and $\phi_G$ in Eq.~(\ref{MG}). The mixing angle between the flavor singlet and octet states is fixed as $\theta=-11^\circ$, and the mixing angle between the pure glueball and the pure heavy quark state is fixed as $\phi_Q=11.6^\circ$. One can check that the physical glueball mass $M_G$ keeps stable within the reasonable ranges of $\phi_Q$ and $\theta$. We adopt $f_q=f_\pi=131$ MeV as an input. Parameter $G_g$ is fixed as $|G_g|=(0.054 \pm 0.008)$ GeV$^3$ from the quenched LQCD calculation. As mentioned earlier, the positive $G_g$ is excluded in our model since it will result in negative values for $M_G^2$. This is consistent with analyses of Ref.~\cite{Cheng:2008ss,Tsai:2011dp}. Actually, within the favored space for all the other parameters the negative values for $G_g$ are always required. Therefore, we fix $G_g=-(0.054 \pm 0.008)$ GeV$^3$ in this analysis and the $\phi_G$ dependence of $M_G$ can be investigated.

As mentioned earlier, the light quark and glueball mixing angle $\phi_G$ has a relatively large variation range, i.e. $\phi_G \in (3,25)^\circ$, we can then investigate the dependence of other quantities on the $\phi_G$ within the range of $\in (3,25)^\circ$. Note that $\phi_G=0$ corresponds to a vanishing mixing between light quarks and glueball. It simply means that as long as the mixing is introduced, the mixing angle will be constrained by other quantities ($G_g$ in this case) and deviate from zero.

\begin{figure}[htb]
\centering
\includegraphics[width=8cm, height=8cm]{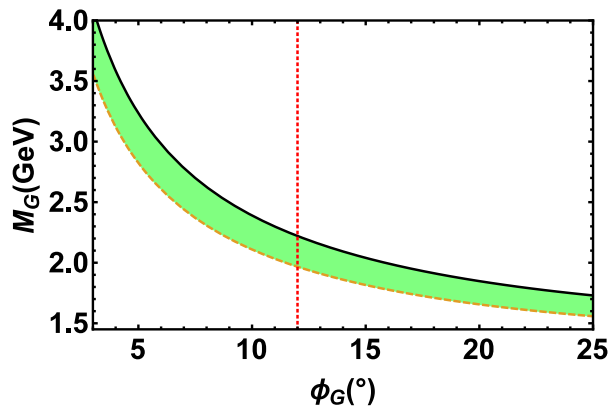}
\caption{The physical glueball mass $M_G$ varies with $\phi_G \in (3 \sim 25)^{\circ}$, with $\theta=-11^\circ$, $\phi_Q=11.6^\circ$, and $f_q=131$ MeV. The solid and dashed line denote the lower and upper limit of $G_g$ and the band in between denotes the uncertainties of the glueball masses for a given $\phi_G$. The vertical dotted line locates the central value of the favored $\phi_G=12^\circ$ from one of the model analyses~\cite{Escribano:2007cd}. }\label{glueball-mass}
\end{figure}

In Fig.~\ref{glueball-mass} we present the physical glueball mass in terms of $\phi_G$ with $G_g= -(0.054 \pm 0.008) \ \gev^3$. It shows that $M_G$ is very sensitive to $\phi_G$. With a small value of $\phi_G=3^\circ$, a large glueball mass of about 3.8 GeV can be extracted. This is even much larger than the pure gauge glueball mass and $\eta_c$ mass. So a small $\phi_G$ like $3^\circ$ is certainly unphysical.  As indicated by the central value of $\phi_G=12^\circ$ from a model analysis~\cite{Escribano:2007cd}, the glueball mass is found to be $M_G \in (2.0,2.2)$ GeV where the uncertainties are given by the uncertainties of $G_g$. This is the range which is not far away from the pure gauge glueball mass by LQCD. It is worth quoting the unquenched LQCD calculations for the pseudoscalar glueball mass in the literature. For instance, the UKQCD Collaboration reported $M_G\simeq 2.5\sim 2.7$ GeV at $m_\pi=280$ and 360 MeV, respectively~\cite{Richards:2010ck}, and $M_G=2.56 \sim 2.60\ \gev$ at $m_\pi=938\sim 650$ MeV were also found by Ref.~\cite{Sun:2017ipk}. Although these results are extracted at relatively high pion mass region, it is very much unlikely that the physical state (a $P$-wave gluonic state) should have a mass lower than that for the scalar glueball (a $S$-wave gluonic state), i.e. around 1.5$\sim$ 1.7 GeV. Our analysis also supports such a scenario. The results in Fig.~\ref{glueball-mass} suggest that low glueball masses, e.g. lower than 1.8 GeV, cannot be accommodated by the mixing mechanism via the axial vector anomaly. As shown by Fig.~\ref{glueball-mass}, even for a much larger and unrealistic value of the mixing angle, the glueball mass will be still higher than 1.5 GeV. This eventually rules out the possibility of a light pseudoscalar glueball around 1.4 GeV.

By substitute Eq.~(\ref{MG}) into the mass density matrix Eq.~(\ref{M-angle}), we obtain the explicit expressions for the mass density matrix elements as follows,
\begin{eqnarray}\label{UMU}
\mathcal{M}_{qsgc}^{11} & = & \frac{M_{\eta}^{2}+M_{\eta^{\prime}}^{2}}{2}
                  +\frac{M_{\eta^{\prime}}^{2}-M_{\eta}^{2}}{12}[2\cos2\theta+\sqrt{2}(\cos\phi_{G}+3)\sec\phi_{G}\sin2\theta]
                  -\frac{2}{\sqrt{3}}\frac{G_{g}}{f_{q}}\sec\phi_{Q}\tan\phi_{G}   \,\nonumber\\
\mathcal{M}_{qsgc}^{12} & = & \mathcal{M}_{qsgc}^{21}=\frac{M_{\eta^{\prime}}^{2}-M_{\eta}^{2}}{6}(2\sqrt{2}\cos2\theta-\cos2\phi_{G}\sec\phi_{G}\sin2\theta)
                    +\frac{2}{\sqrt{6}}\frac{G_{g}}{f_{q}}\sec\phi_{Q}\tan\phi_{G} \, \nonumber\\
\mathcal{M}_{qsgc}^{13} & = & \mathcal{M}_{qsgc}^{31}=\frac{\sqrt{2}G_{g}}{f_{q}} \, \nonumber\\
\mathcal{M}_{qsgc}^{14} & = & \mathcal{M}_{qsgc}^{41}=\frac{\sqrt{2}G_{g}}{f_{q}}\tan\phi_{Q} \, \nonumber\\
\mathcal{M}_{qsgc}^{22} & = & \frac{M_{\eta}^{2}+M_{\eta^{\prime}}^{2}}{2}-\frac{M_{\eta^{\prime}}^{2}-M_{\eta}^{2}}{24}[4\cos2\theta
                   +\sqrt{2}(5\cos2\phi_{G}+3)\sec\phi_{G}\sin2\theta]-\frac{G_{g}}{\sqrt{3}f_{q}}\sec\phi_{Q}\tan\phi_{G} \, \nonumber\\
\mathcal{M}_{qsgc}^{23} & = & \mathcal{M}_{qsgc}^{32}=\frac{G_{g}}{f_{q}}-\sqrt{\frac{3}{2}}(M_{\eta^{\prime}}^{2}
                   -M_{\eta}^{2})\cos\theta\cos\phi_{Q}\sin\theta\sin\phi_{G} \, \nonumber\\
\mathcal{M}_{qsgc}^{24} & = & \mathcal{M}_{qsgc}^{42}=\frac{G_{g}}{f_{q}}\tan\phi_{Q}-\sqrt{\frac{3}{2}}(M_{\eta^{\prime}}^{2}
                  -M_{\eta}^{2})\cos\theta\sin\phi_{Q}\sin\theta\sin\phi_{G} \, \nonumber\\
\mathcal{M}_{qsgc}^{33} & = & (M_{\eta}^{2}\sin^{2}\theta+M_{\eta^{\prime}}^{2}\cos^{2}\theta+\frac{M_{\eta^{\prime}}^{2}
                   -M_{\eta}^{2}}{\sqrt{2}}\cos\phi_{G}\sin\theta\cos\theta)\cos^{2}\phi_{Q}  -\frac{\sqrt{3}G_{g}}{f_{q}}\cos\phi_{Q}\cot\phi_{G}+M_{\eta_{c}}^{2}\sin^{2}\phi_{Q} \, \nonumber\\
\mathcal{M}_{qsgc}^{34} & = & \mathcal{M}_{qsgc}^{43}=[\frac{M_{\eta}^{2}+M_{\eta^{\prime}}^{2}-2M_{\eta_{c}}^{2}}{4}
                   -\frac{\sqrt{3}G_{g}}{2f_{q}}\sec\phi_{Q}\cot\phi_{G}   +\frac{M_{\eta^{\prime}}^{2}-M_{\eta}^{2}}{4}(\cos2\theta+\sqrt{2}\cos\theta\cos\phi_{G}\sin\theta)]\sin2\phi_{Q} \, \nonumber\\
\mathcal{M}_{qsgc}^{44} & = & M_{\eta_{c}}^{2}\cos^{2}\phi_{Q}+\frac{\sin^{2}\phi_{Q}}{2}[(M_{\eta}^{2}+M_{\eta^{\prime}}^{2}) +(M_{\eta^{\prime}}^{2}-M_{\eta}^{2})(\cos2\theta+\sqrt{2}\cos\theta\sin\theta\cos\phi_{G})]\nonumber\\
&& -\frac{\sqrt{3}G_{g}}{f_{q}}\tan\phi_{Q}\cot\phi_{G}\sin\phi_{Q} \ .
\end{eqnarray}
There are apparent features arising from the mixings described by the above equation array. In the light flavor sector the mixing between $|q\bar q\rangle$ and $ |s \bar s\rangle$ is dominated by the flavor singlet and octet mixing as expected. The glueball mixing contributions are at order of $G_g/f_q$ but it enters into the light quark submatrix with a suppression factor $\tan\phi_G$. Due to the small value of $\phi_G$, one would not expect significant contributions from the glueball mixings in $\eta$ and $\eta'$ as demonstrated by many studies. The mixing between the heavy and light flavors can be seen in $\mathcal{M}_{qsgc}^{14,24}$, which is at order of $\tan\phi_Q$ and can be neglected. This is anticipated due to the large mass difference between $\eta_c$ and $\eta \ (\eta^\prime)$. The mixing between glueball and heavy flavor $c\bar{c}$ can be seen from $\mathcal{M}_{qsgc}^{34}$, where cancellations among the terms are present. Furthermore, this element is proportional to $\sin 2\phi_Q$. Given the small value of $\phi_Q$, this factor also imposes a suppression to the mixing effects.

In the limit of small values for $\phi_G$ and $\phi_Q$, the element $\mathcal{M}_{qsgc}^{33}$ can be approximated as
\begin{eqnarray}
\mathcal{M}_{qsgc}^{33}&\approx & -\frac{\sqrt{3}G_g/f_q}{\sin\phi_G}+M_{\eta'}^2-(M_{\eta'}^2-M_\eta^2)(\sin^2\theta-\frac{1}{\sqrt{2}}\sin\theta\cos\theta)+M_{\eta_c}^2\sin^2\phi_Q \,\nonumber\\
&=& -\frac{\sqrt{3}G_g/f_q}{\sin\phi_G}+M_{\eta'}^2 + (M_{\eta'}^2-M_\eta^2)\sin\theta\cos(\theta+\theta_i) {1 \over \sin\theta_i} +M_{\eta_c}^2\sin^2\phi_Q \,\nonumber\\
&\approx& M_G^2 +M_{\eta_c}^2\sin^2\phi_Q
\ ,
\end{eqnarray}
where we keep the correction from $\eta_c$ to show the suppressed contributions from the heavy flavor part. The last line is obtained by substituting Eq.~(\ref{MG-1}) into the equation with $\sin\theta_i=\sqrt{2/3}$. It is interesting to compare the above expression with Eq.~(\ref{MG-1}). It shows that the mixing effects on the mass of glueball from the quark states are indeed suppressed. Apart from the term of $M_{\eta_c}^2\sin^2\phi_Q$ from the heavy flavor mixing, corrections from the light flavor singlet and octet mixings will introduce cancellations. The numerical calculation indeed suggests that such mixing effects cannot significantly change the pure glueball mass. Alternatively, it implies that the physical $\eta_c$ will have subleading mixing contributions from the glueball. This feature can be seen by the element $\mathcal{M}_{qsgc}^{44}$, and is consistent with the experimental observations. A detailed investigation of this aspect has been presented in Ref.~\cite{Tsai:2011dp}.

Combining the mass density matrix elements in Eq.~(\ref{UMU}) with those defined in Eq.~(\ref{Mdynexp}), all the unknown parameters can be written in terms of $\phi_G$ and the constrained parameters as follows,
\begin{eqnarray}\label{parameters}
M_G &=& \large\left(\right. -\frac{1}{\cos\phi_G\sin\theta_i \cos\phi_Q}\{\frac{\sqrt2 G_g/f_q}{\sin\phi_G}-[-M_\eta^2(\cos\theta \cos\theta_i-\sin\theta \cos\phi_G \sin\theta_i)
\sin\theta  \cos\phi_Q      \,\nonumber\\
& &+M_{\eta'}^2(\sin\theta \cos\theta_i + \cos\theta \cos\phi_G \sin\theta_i)
\cos\theta \cos\phi_Q]\} \left. \large\right)^{1 \over 2} \ ,\nonumber\\
G_q   &=&  {(M_{\eta^\prime}^2 - M_\eta^2 ) f_s \over 2}  ( \sin2\theta \cos2\theta_i \cos\phi_G +
            \sin2\theta cos^2\theta_i \sin\phi_G \tan\phi_G + \cos2\theta \sin2\theta_i )
            -  G_g { \sqrt2 f_s \over f_q } \cos\theta_i \tan\phi_G \sec\phi_Q   \ ,\nonumber\\
G_s   &=& (M^2_{\eta^\prime}-M_\eta^2) {f_q \over 2\sqrt2} ( \cos2\theta_i \cos\phi_G \sin2\theta + \cos2\theta \sin2\theta_i + \cos^2\theta_i \sin2\theta \sin\phi_G \tan\phi_G    ) - G_g \cos\theta_i \sec\phi_Q \tan\phi_G \,\nonumber\\
G_c   &=& -{\sqrt2 f_c \over f_q} G_g \csc\theta_i \cot\phi_G \sin\phi_Q \tan\phi_Q +  (M_{\eta_c}^2 \cos^2\phi_Q - m_{cc}^2) f_c \nonumber\\
             &\ &  +
           [(M_{\eta^\prime}^2 - M_\eta^2)  \sin\theta \cos\theta \cot\theta_i \cos\phi_G + ( M_\eta^2 \sin^2\theta + M_{\eta^\prime}^2 \cos^2\theta ) ]f_c \sin^2\phi_Q    \nonumber\\
         \ ,\nonumber\\
m_{qq}^2&=& ( M_\eta^2 - M_{\eta^\prime}^2 ) [ { f_s \over \sqrt2 f_q } ( \sin2\theta \cos2\theta_i \cos\phi_G
             + \sin2\theta \cos^2\theta_i \sin\phi_G \tan\phi_G +\cos2\theta \sin2\theta_i )
              \nonumber\\
             &\ &
             - {1 \over 8} (3 + \cos2\phi_G) \sin2\theta \sin2\theta_i \sec\phi_G
             + { 1 \over 2} \cos^2\theta \cos2\theta_i ]
             + \cos^2\theta { M_{\eta^\prime}^2 +  M_\eta^2 \over 2 }
             + \sin^2\theta ( M_\eta^2 \sin^2\theta_i + M_{\eta^\prime}^2  \cos^2\theta_i ) \nonumber\\
             &\ &+ G_g {2 \over f_q} \tan\phi_G \sec\phi_Q \cos\theta_i ( {  f_s \over f_q }  - 1  )
                 \ ,\nonumber\\
m_{ss}^2&=&   ( M_\eta^2 - M_{\eta^\prime}^2 ) [ {f_q \over 2\sqrt2 f_s} ( \sin2\theta \cos2\theta_i \cos\phi_G + \sin2\theta \cos^2\theta_i \sin\phi_G \tan\phi_G + \cos2\theta \sin2\theta_i ) + \sin2\theta \sin\theta_i \cos\theta_i \cos\phi_G \nonumber\\
        &\ &  - \sin\theta \cos\theta \cos^2\theta_i \cot\theta_i \sin\phi_G \tan\phi_G   ]
        + \sin^2\theta ( M_\eta^2 \cos^2\theta_i + M_{\eta^\prime}^2 \sin^2\theta_i  )
        + \cos^2\theta ( M_\eta^2 \sin^2\theta_i + M_{\eta^\prime}^2 \cos^2\theta_i )
        \nonumber\\
        &\ &+ G_g \cos\theta_i \tan\phi_G \sec\phi_Q  ( {1 \over f_s}  - {1 \over f_q } )
         \ ,\nonumber\\
m_{sg}^2&=& G_g ( \frac{\sqrt{2} \cot\theta_i}{f_q } -\frac{1}{f_s} ) + ( M_\eta^2 - M_{\eta^\prime}^2 ) \sin\theta \cos\theta \csc\theta_i
             \sin\phi_G \cos\phi_Q    \nonumber\\
        &=& G_g ( \frac{1}{f_q } -\frac{1}{f_s} ) + ( M_\eta^2 - M_{\eta^\prime}^2 ) \sin\theta \cos\theta \csc\theta_i \sin\phi_G \cos\phi_Q
              \ ,\nonumber\\
m_{cg}^2&=& - G_g (\frac{1}{f_c} + \frac{1}{f_q} \sqrt{2} \csc\theta_i \cot\phi_G \sin\phi_Q  )
           + [( M_{\eta^{\prime}}^2 - M_\eta^2) (\sin2\theta \cot\theta_i \cos\phi_G  + \cos2\theta  )
         +  ( M_{\eta^{\prime}}^2 + M_\eta^2) - 2  M_{\eta_c}^2 ] {\sin2\phi_Q  \over 4}
         \ ,\nonumber\\
m_{qc}^2&=&  (m_{cc}^2 - M_{\eta_c}^2 \cos^2\phi_Q){\sqrt2 f_c \over f_q}
            + {\sqrt2 f_c \over f_q} \sin^2\phi_Q  [( M_\eta^2 - M_{\eta^\prime}^2 )  \cos\theta \cos\phi_G \cot\theta_i \sin\theta
            \nonumber\\
            &\ &
            -   ( M_\eta^2 \sin^2\theta +  M_{\eta^\prime}^2 \cos^2\theta  )]
              +  {\sqrt2 G_g \over f_q} \tan\phi_Q ( 1 +{\sqrt2 f_c \over f_q} \cot\phi_G \csc\theta_i \sin\phi_Q  )   \ ,\nonumber\\
m_{sc}^2&=&  (m_{cc}^2 - M_{\eta_c}^2 \cos^2\phi_Q) {f_c \over f_s} + (M_\eta^2- M_{\eta^\prime}^2) \cos\theta \sin\theta \sin\phi_Q  [ \csc\theta_i \sin\phi_G + {f_c \over f_s} \cos\phi_G \cot\theta_i  \sin\phi_Q  ] \nonumber\\
        &\ &
           - {f_c \over f_s } \sin^2\phi_Q  ( M_\eta^2 \sin^2\theta + M_{\eta^\prime}^2 \cos^2\theta ) + {\sqrt2 G_g \over f_q } \tan\phi_Q( \cot\theta_i + {f_c \over f_s} \cot\phi_G \csc\theta_i \sin\phi_Q  )
\ ,\nonumber\\
m_{cq}^2&=&  ( M_\eta^2 - M_{\eta^\prime}^2 )  { f_s \over 2 f_c }
             ( \sin2\theta \cos2\theta_i \cos\phi_G +\sin2\theta \cos^2\theta_i \sin\phi_G \tan\phi_G + \cos2\theta \sin2\theta_i )\nonumber\\
              &\ & + G_g {\sqrt2 \over f_q} ( {f_s \over f_c   }  \cos\theta_i \tan\phi_G \sec\phi_Q  + \tan\phi_Q  )     \ ,\nonumber\\
m_{cs}^2&=&  ( M_\eta^2 - M_{\eta^\prime}^2 ) { f_q \over 2\sqrt2 f_c } ( \sin2\theta \cos2\theta_i \cos\phi_G + \sin2\theta \cos^2\theta_i \sin\phi_G \tan\phi_G + \cos2\theta \sin2\theta_i \nonumber\\
        &\ & + \sin\theta \cos\theta \csc\theta_i \sin\phi_G\sin\phi_Q ) + G_g ( {1 \over f_c} \cos\theta_i \tan\phi_G \sec\phi_Q + { 1 \over f_q }  \tan\phi_Q ) \ .
\end{eqnarray}

One notices that the elements between the light and heavy flavor mixings are suppressed explicitly either by the mixing angle or a term of $(M_{\eta_c}^2\cos^2\phi_Q-m_{cc}^2)$. This is understandable due to the large mass differences between $\eta_c$ and $\eta \ (\eta')$.

To see more clearly the dependence of the mixing matrix elements on the mixing angles and decay constants, we substitute the values of those fixed parameters, i.e.  $M_\eta, \ M_{\eta^\prime}, \ M_{\eta_c}, \ \theta_i$, and $G_g $, into the above equations. The mixing elements will be explicit functions of $\theta, \ \phi_G, \ \phi_Q$ and the decay constants. We can then investigate their relations by numerical calculations.
\begin{eqnarray}\label{parameters-2}
M_G &=&  [( -{1.8G_g \over f_q}  \csc\phi_G-0.082) \sec\phi_G+0.90]^{1/2} \ ,\nonumber\\
G_q   &=& f_s [ 0.29 \cos2\theta - {0.83G_g \over f_q} \tan\phi_G - 0.10 \sin2\theta \cos\phi_G ( 1 - \tan^2\phi_G )  ]    \ ,\nonumber\\
G_s   &=& f_q [ 0.21 \cos2\theta - 0.072 \sin2\theta \cos\phi_G + 0.073 \sin2\theta \sin\phi_G \tan\phi_G ]
           - 0.59 G_g \tan\phi_G \, \nonumber\\
G_c   &=& [(8.9 \cos^2\phi_Q - m_{cc}^2)  + \sin^2\phi_Q ( 0.90-0.08\cos\phi_G ) - \sin\phi_Q \tan\phi_Q \cot\phi_G {\sqrt 3 G_g \over f_q} ] f_c \ ,\nonumber\\
m_{qq}^2&=& [ 0.71\cos^2\theta +0.51\sin^2\theta + \sin2\theta \sec\phi_G ( 0.22+0.073 \cos2\phi_G)
          - {1.2G_g \over f_q} \tan\phi_G \sec\phi_Q (1- {f_s \over f_q}) ]  \nonumber\\
          &\ &  - {f_s \over f_q} (0.41\cos2\theta - 0.14 \sin2\theta\cos\phi_G(1 - \tan^2\phi_G))
        \ ,\nonumber\\
m_{ss}^2&=&0.51 \cos^2\theta -  \sin\theta \cos\theta \cos\phi_G ( 0.58 - 0.15 \tan^2\phi_G ) + 0.71 \sin^2\theta  \nonumber\\
          &\ & - {0.59 G_g\tan\phi_G \over f_s} (  {f_s \over f_q} - 1 )
          + {f_q \over f_s} [ 0.072 \sin2\theta \cos\phi_G ( 1 - \tan^2\phi_G ) - 0.21 \cos2\theta ]    \ ,\nonumber\\
m_{sg}^2&=& \frac{G_g}{f_s} ( { f_s \over f_q } - 1 ) -  0.74 \sin\theta \cos\theta \sin \phi_G \ ,\nonumber\\
m_{cg}^2&=& -\frac{G_g}{f_c} ( 1 + \sqrt 3 { f_c \over f_q } \cot\phi_G \sin\phi_Q )
             -   ( 0.082 \cos\phi_G + 8.0 )  \sin\phi_Q \cos\phi_Q \ ,\nonumber\\
m_{qc}^2&=& [\sqrt 2 (m_{cc}^2 - 8.9\cos^2\phi_Q) + \sin^2\phi_Q ( 0.12 \cos\phi_G-1.27 + { 2.45G_g \over f_q  } \cot\phi_G
           \sec\phi_Q
              )
             ] \frac{f_c}{f_q} + {\sqrt 2 G_g\tan\phi_Q \over f_q } \ ,\nonumber\\
m_{sc}^2&=&( m_{cc}^2 - 8.9 \cos^2\phi_Q + {\sqrt 3 G_g \over f_q} \cot\phi_G \sin\phi_Q \tan\phi_Q - \sin^2\phi_Q ( 0.90-0.082\cos\phi_G
              )   )   \frac{f_c} {f_s}  \nonumber\\
              &\ &
              + G_g {\tan\phi_Q \over f_q} + 0.14 \sin\phi_G \sin\phi_Q \ ,\nonumber\\
m_{cq}^2&=&   - \frac{f_s}{f_c} [ 0.29\cos2\theta - {0.82G_g \over f_q} \tan\phi_G \sec\phi_Q - {\sqrt2 G_g \over f_q} {f_c \over
               f_s}
              \tan\phi_Q  - 0.10\sin2\theta\cos\phi_G ( 1 - \tan^2\phi_G)  ] \ ,\nonumber\\
m_{cs}^2&=& -{f_q \over f_c} [ 0.19 + 0.027 \cos\phi_G( 1 - \tan^2\phi_G )  ] + {\sec\phi_Q \over f_c} ( 0.58G_g \tan\phi_G + G_g
             {f_c \over f_q}  \sin\phi_Q ) + 0.14\sin\phi_G \sin\phi_Q  \ .
\end{eqnarray}
In the above equation array $M_G$ shows explicit dependence on $G_g$ and $\phi_G$. $G_q$, $G_s$ and $G_c$ are proportional to $f_s$, $f_q$ and $f_c$, respectively. One notices that $G_c, \ m_{qc}^2$, and  $m_{sc}^2$ contains the large cancellation term $(M_{\eta_c}^2 \cos^2\phi_Q - m_{cc}^2)$. Thus, $G_c$ turns out to be sensitive to $m_{cc}^2$ and $\phi_Q$. These quantities are also dependent on $\phi_G$ due to the factor $\cot\phi_G$ there. There are also large cancellations in $m^2_{qq}$ which is sensitive to both  ${f_s/f_q}$ and $\theta$. In contrast, the cancellation in $m_{ss}^2$ is relatively small, and it shows small dependence on ${f_s/f_q}$. $m_{sg}^2$ shows sensitivities to ${f_s/f_q}$ since it contains an SU(3) flavor symmetry breaking factor $({f_s/f_q}-1)$. A cancellation also occurs in $m^2_{cg}$, and $m^2_{cg}$ increases with the decreasing $\phi_G$. In contrast, it decreases when the $\phi_Q$ gets smaller. In Eq.~(\ref{parameters-2}), we have also listed $m_{cq}^2$ and $m_{cs}^2$ in terms of $\phi_Q, \ m_{cc}^2$ and $\phi_G$ in comparison with $m_{qc}^2$ and $m_{sc}^2$. However, they show little dependence on all these mixing angles.

By adopting $G_g=-0.054$ GeV$^3$, $\phi_Q=11.6^\circ$, $\phi_G=12^\circ$, $f_c=487.4$ MeV, $f_q=131$ MeV, and taking the ratio $f_s/f_q=1.2$ and $1.3$ in order to examine the sensitivity of the SU(3) flavor symmetry breaking, we can determine all the other quantities and they are listed in Table~\ref{parameters-value} for the two ratios of $f_s/f_q$, respectively. One notices that some of these parameters do not explicitly depend on $f_s$ as discussed above. Thus, they do not change values when taking different ratios for $f_s/f_q$.

\begin{table}[htb]
\caption{The numerical values of all the parameters with $G_g=-0.054$ GeV$^3$ and $\phi_G=12^\circ$ fixed. The two quantities, $m_{qc}^{2*}$ and $m_{sc}^{2*}$ involve more complicated issues and are sensitive to $m_{cc}^2$ and $\phi_G$. Further detailed discussions can be found in the context.\label{parameters-value}}
\begin{tabular*}{\textwidth}{@{\extracolsep{\fill}}c c c c c c c c c c c c c c c c}
  \hline
  \hline
  $f_s/f_q$ & $M_G$(GeV) & $m_{qq}^2$(GeV)$^2$  & $m_{ss}^2$ & $m_{sg}^2$& $m_{cg}^2$& $m_{qc}^{2*}$ & $m_{sc}^{2*}$& $m_{cq}^2$& $m_{cs}^2$& $G_q$(GeV)$^3$& $G_s$& $G_c$\\
  \hline
   1.2 & 2.1 & 0.055  & 0.45 & $-0.041$ & $-0.81$ & 0.87   &  0.50  &  $-0.24$  &   $-0.15$  &  0.060  &  0.035  &  $-0.092$ \\
  \hline
   1.3 & 2.1 & 0.0012 & 0.47 & $-0.067$ & $-0.81$ & 0.87   &  0.46  &  $-0.25$  &   $-0.15$  &  0.065  &  0.035  &  $-0.092$\\
  \hline \hline
\end{tabular*}
\end{table}

The correlations among the parameters should be further discussed. In Table~\ref{parameters-value} it shows that $m_{qq}^2$ is quite sensitive to the SU(3) flavor symmetry breaking ratio $f_s/f_q$. Namely, $m_{qq}^2$ changes order with the ratio $f_s/f_q$ varying from $1.2$ to $1.3$. In contrast,  $m_{sg}^2$ changes by about a factor of 1.5. Such a dependence can be seen from Eq.~(\ref{parameters}). Taking the small angle limit for $\phi_G$ and $\theta$, $m_{qq}^2$ can be approximated by
\begin{equation}
m_{qq}^2\simeq 0.71 \cos^2\theta - (0.41-0.14\sin2\theta){f_s \over f_q} + 0.29 \sin2\theta \ ,
\end{equation}
where significant cancellations occur with the increasing ratio of $f_s/f_q$. With the dominance of the constant term the value of $m_{qq}^2$  will decrease due to a cancellation caused by the increasing ratio of $f_s/f_q$. Similar phenomenon happens to $m_{sg}^2$. It is reasonable that $f_q$ and $f_s$ would affect the light quark mass term that are related to the light flavor states $\eta_q$ and $\eta_s$. Since the other parameters keep stable with the varying ratio of $f_s/f_q$ in a reasonable range, we will focus on the results with $f_s/f_q=1.2$ in the following discussions.

\subsection{Extracting topological susceptibilities for pseudoscalar mesons}

It is noticeable that the anomaly matrix elements $G_q$, $G_s$ and $G_c$ are of the same order of $G_g$. $G_q$ and $G_s$ are quite large which is an indication of the important role played by the anomaly term in the U(1) Goldstone boson~\cite{Witten:1979vv,Veneziano:1979ec}. As also discussed in Ref.~\cite{Cheng:2008ss}, the anomaly terms $\langle 0|\alpha_sG{\tilde G}/(4\pi)|\eta\rangle$ and $\langle 0|\alpha_sG{\tilde G}/(4\pi)|\eta^{\prime} \rangle$ could be related to the topological susceptibility. In our scheme we obtain,
\begin{eqnarray}\label{suscept}
 \langle 0|\alpha_sG{\tilde G}/(4\pi)|\eta \rangle &=& 0.016 \ \gev^3, \nonumber \\
 \langle 0|\alpha_sG{\tilde G}/(4\pi)|\eta^{\prime} \rangle &=& 0.051 \ \gev^3, \nonumber \\
 \langle 0|\alpha_sG{\tilde G}/(4\pi)|G \rangle &=&-0.084 \ \gev^3,  \nonumber \\
  \langle 0|\alpha_sG{\tilde G}/(4\pi)|\eta_{c} \rangle &=& -0.079 \ \gev^3,
\end{eqnarray}
where the central values of $\phi_G=12^\circ$ and $m_{cc}^2=M_{\eta_c}^2$ are adopted. These results are consistent with the LQCD calculations, namely, $\langle 0|\alpha_sG{\tilde G}/(4\pi)|\eta \rangle\approx 0.021$ GeV$^3$~\cite{Novikov:1979uy}, $\langle 0|\alpha_sG{\tilde G}/(4\pi)|\eta^{\prime} \rangle\approx  0.035$ GeV$^3$~\cite{DelDebbio:2004ns}, which have been determined in the chiral limit on LQCD, and $G_g=-(0.054\pm 0.008)\gev ^3$ calculated in the quenched approximation~\cite{Chen:2005mg}.

\begin{figure}[htb]
\centering
\subfigure{\includegraphics[width=5cm, height=4cm]{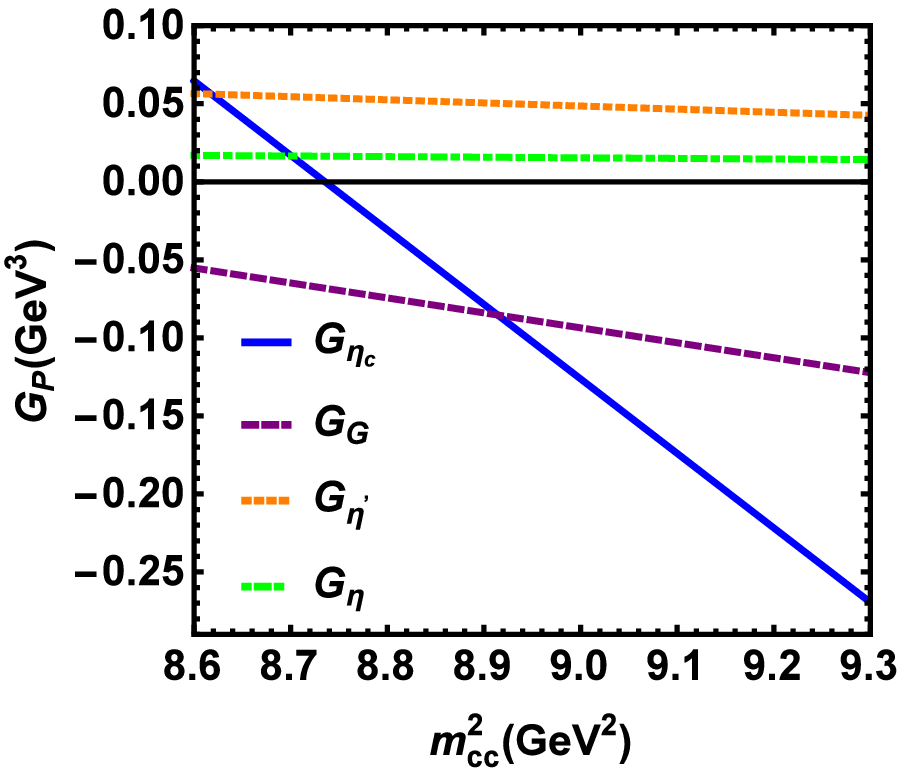}}\hspace{0.5cm}\subfigure{\includegraphics[width=5cm, height=4cm]{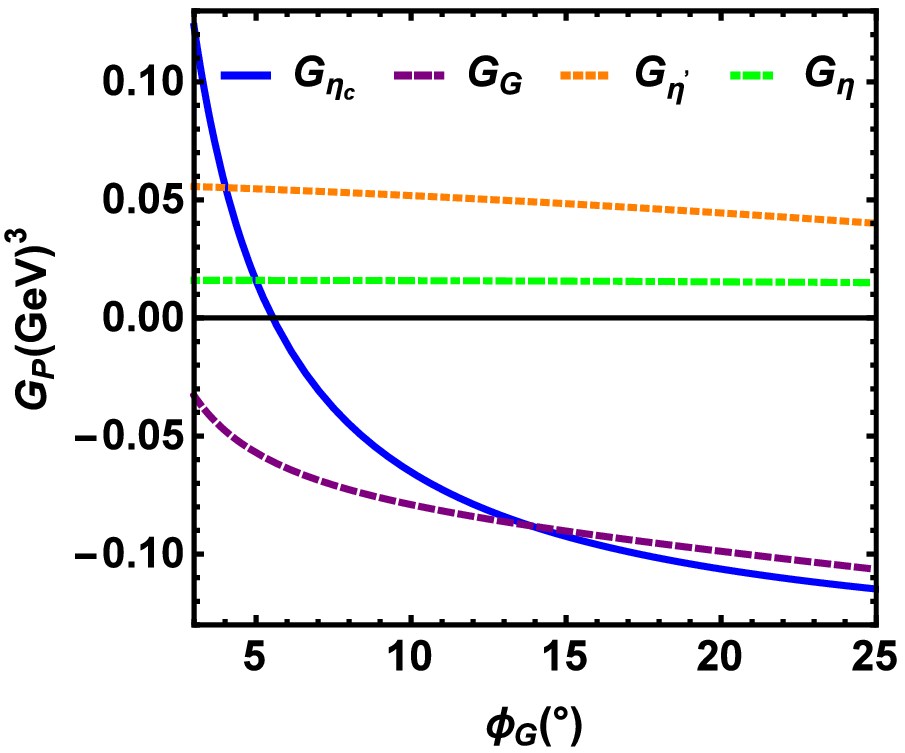}}\hspace{0.5cm}\subfigure{\includegraphics[width=5cm, height=4cm]{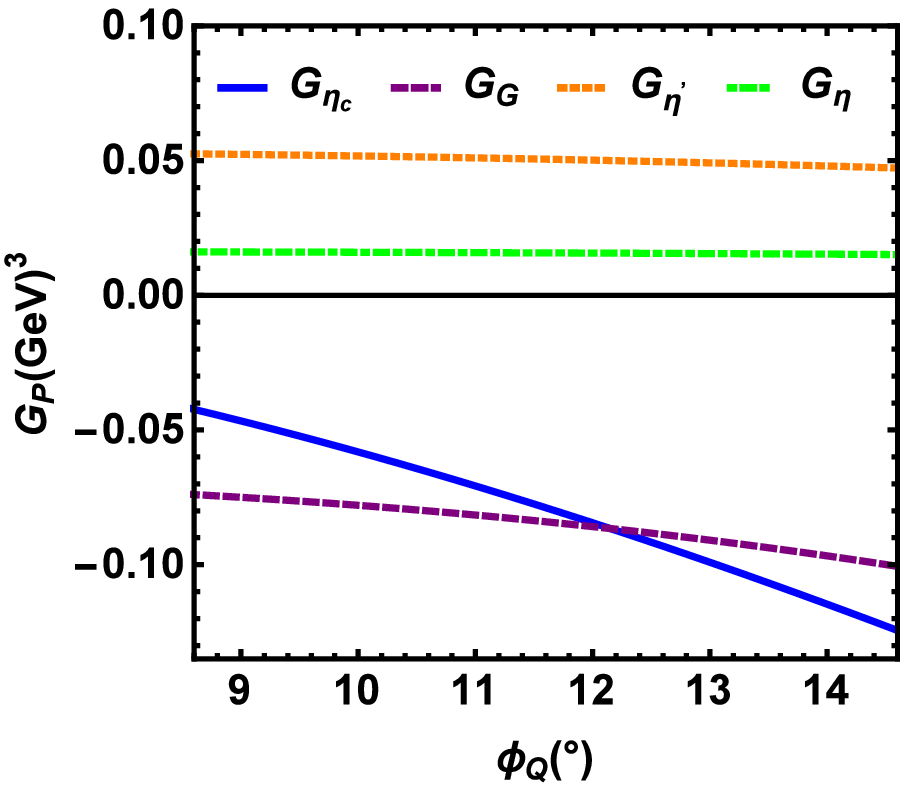}}
 \vspace{0cm}
\caption{The dependence of $G_P$ on $m_{cc}^2$, $\phi_G$ and $\phi_Q$. The results shown from left to right panels are extracted with (left) $\phi_G=12^\circ$ and $\phi_Q=11.6^\circ$,  (middle) $m_{cc}^2=M_{\eta_c}^2$ and $\phi_Q=11.6^\circ$, and (right)  $\phi_G=12^\circ$ and $m_{cc}^2=M_{\eta_c}^2$, respectively.}\label{Gphys}
\end{figure}

In order to estimate the uncertainties arising from the parameter ranges, we plot the topological susceptibility $G_P\equiv\langle 0|\alpha_sG{\tilde G}/(4\pi)|P \rangle$, with $P$ stands for the physical states $\eta$, $\eta^{\prime}$, $G$ and $\eta_{c}$, in terms of $m_{cc}^2,\phi_Q$ and $\phi_G$ in Fig.~\ref{Gphys}. We consider the dependence of $G_P$ on these three quantities in three cases, i.e. (a) $G_P$ dependence on $m_{cc}$ (with $\phi_G=12^\circ$, $\phi_Q=11.6^\circ$ fixed); (b) $G_P$ dependence on $\phi_G$ (with $m_{cc}^2=M_{\eta_c}^2$, $\phi_Q=11.6^\circ$ fixed); and (c) $G_P$ dependence on $\phi_Q$ (with $\phi_G=12^\circ$, $m_{cc}^2=M_{\eta_c}^2$ fixed). It shows that $G_\eta$ and $G_{\eta'}$ are not sensitive  to $m_{cc}$, $\phi_G$ and $\phi_Q$ mainly because the mixing between $\eta_{q,s}$ and $\eta_Q$ are small. More significant sensitivities of $G_G$ and $G_{\eta_c}$ indicate the non-negligible effects arising from the mixing between $|\eta_Q\rangle$ and $|g\rangle$.

To be specific, in Fig.~\ref{Gphys} (a) $G_{\eta_c}$ is the only one sensitive to the value of $m_{cc}^2$ due to the presence of the dominant cancellation term $(8.9 \cos^2\phi_Q - m_{cc}^2)$ as shown in Eq.~(\ref{parameters-2}). By adopting $G_c=\langle 0|\alpha_sG{\tilde G}/(4\pi)|\eta_{c} \rangle = -0.079 \ \gev^3$, the corresponding value of $m_{cc}^2)$ becomes close to $M_{\eta_c}^2$ as expected. In contrast, in Fig.~\ref{Gphys} (b) when fixing $m_{cc}^2=M_{\eta_c}^2$ and $\phi_Q=11.6^\circ$ the $\phi_G$ dependence of both $G_{\eta_c}$ and $G_G$ are sensitive in the small value range of $\phi_G\lesssim 10^\circ$. This suggests that $\phi_G$ can be well constrained by the mixing angle in our scenario. In Fig.~\ref{Gphys} (c), by fixing $\phi_G=12^\circ$ and $m_{cc}^2=M_{\eta_c}^2$, all the quantities appear to be stable in terms of $\phi_Q$ in a relatively broad range. The favored value $\phi_Q=11.6^\circ$ corresponds to an overall reasonably good description of all the other quantities.

\begin{figure}[htb]
\centering
\subfigure{\includegraphics[width=5cm, height=4cm]{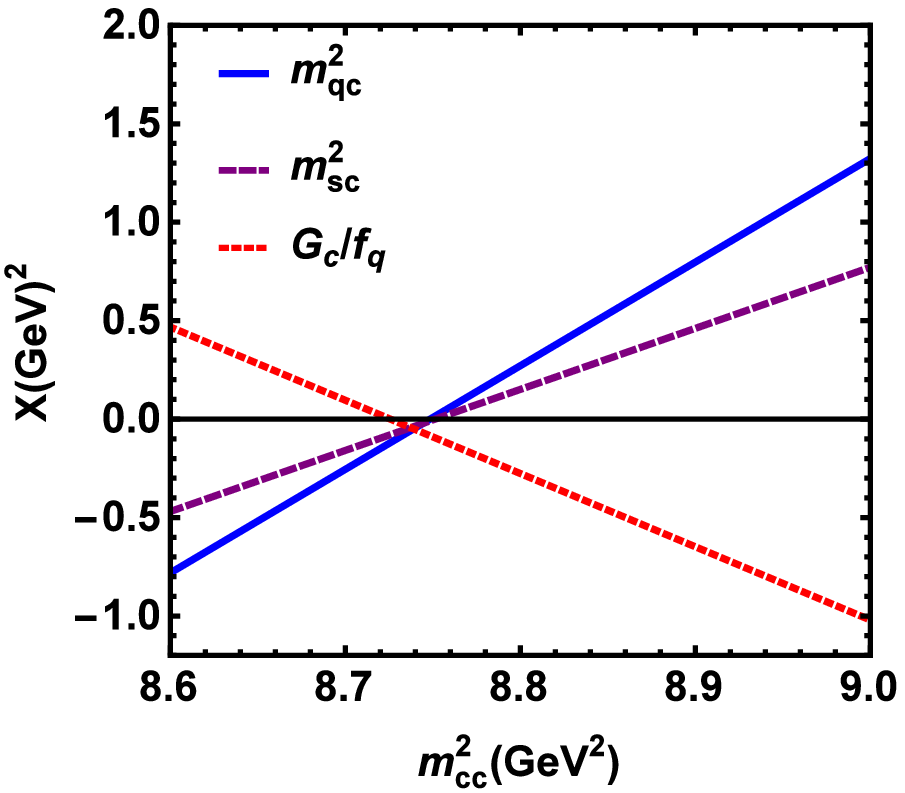}}\hspace{0.5cm}\subfigure{\includegraphics[width=5cm, height=4cm]{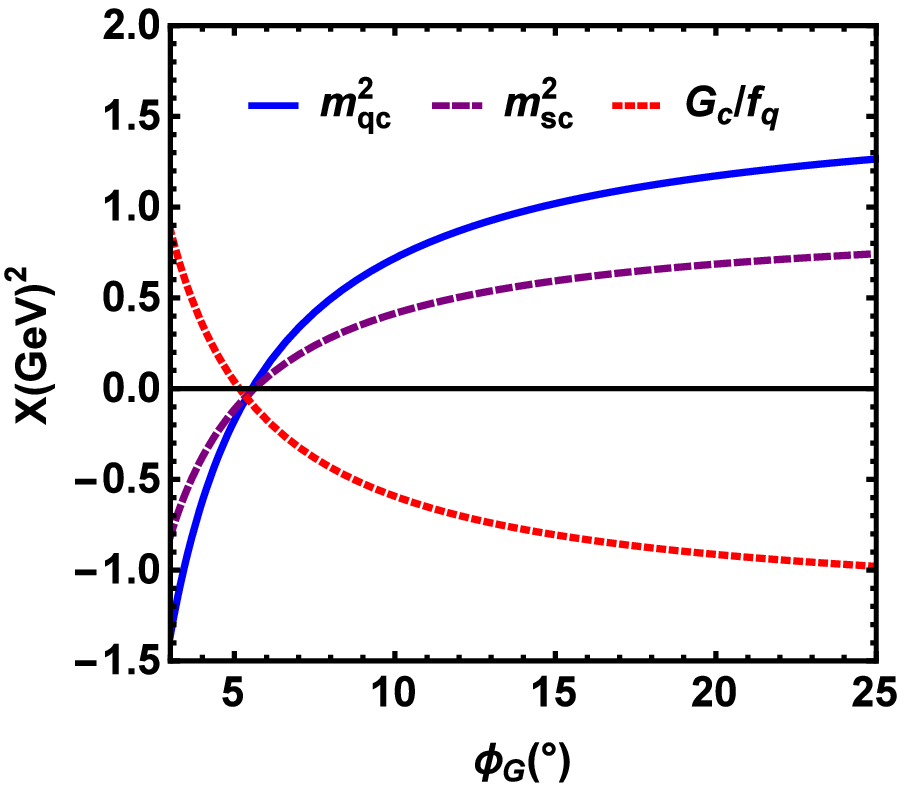}}\hspace{0.5cm}\subfigure{\includegraphics[width=5cm, height=4cm]{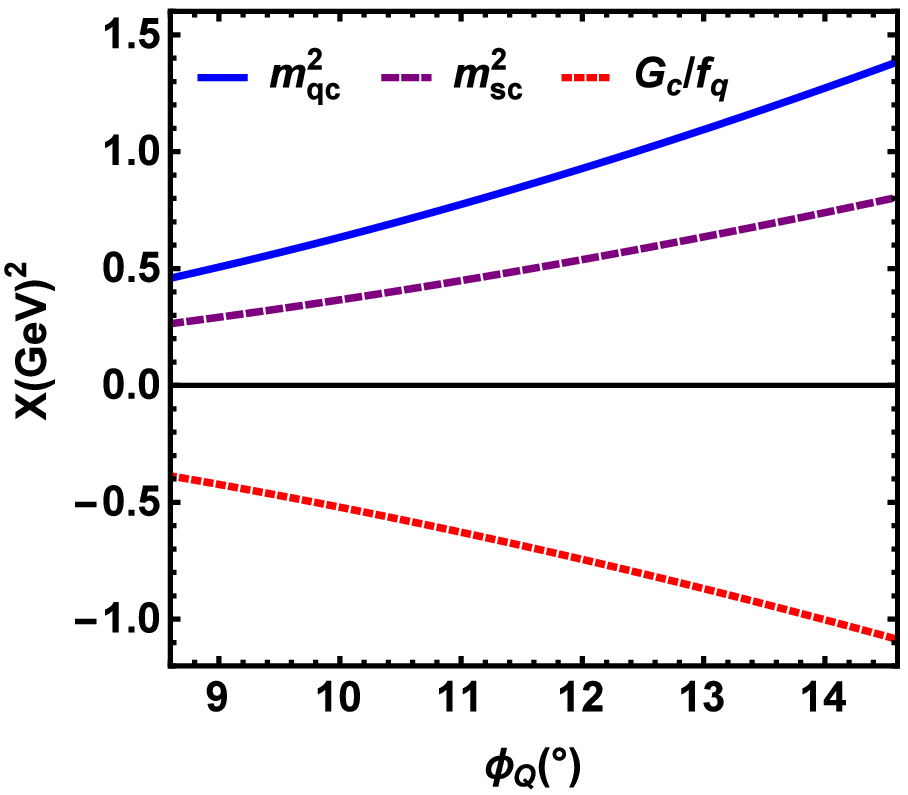}}
 \vspace{0cm}
\caption{The dependence of $m_{qc}^2$, $m_{sc}^2$ and $G_c/f_q$ on $m_{cc}^2$, $\phi_G$ and $\phi_Q$. The results shown from left to right panels are extracted with (left) $\phi_G=12^\circ$ and $\phi_Q=11.6^\circ$,  (middle) $m_{cc}^2=M_{\eta_c}^2$ and $\phi_Q=11.6^\circ$, and (right)  $\phi_G=12^\circ$ and $m_{cc}^2=M_{\eta_c}^2$, respectively.}\label{mqc-msc}
\end{figure}

\begin{figure}[htb]
\centering
\subfigure{\includegraphics[width=5cm, height=4cm]{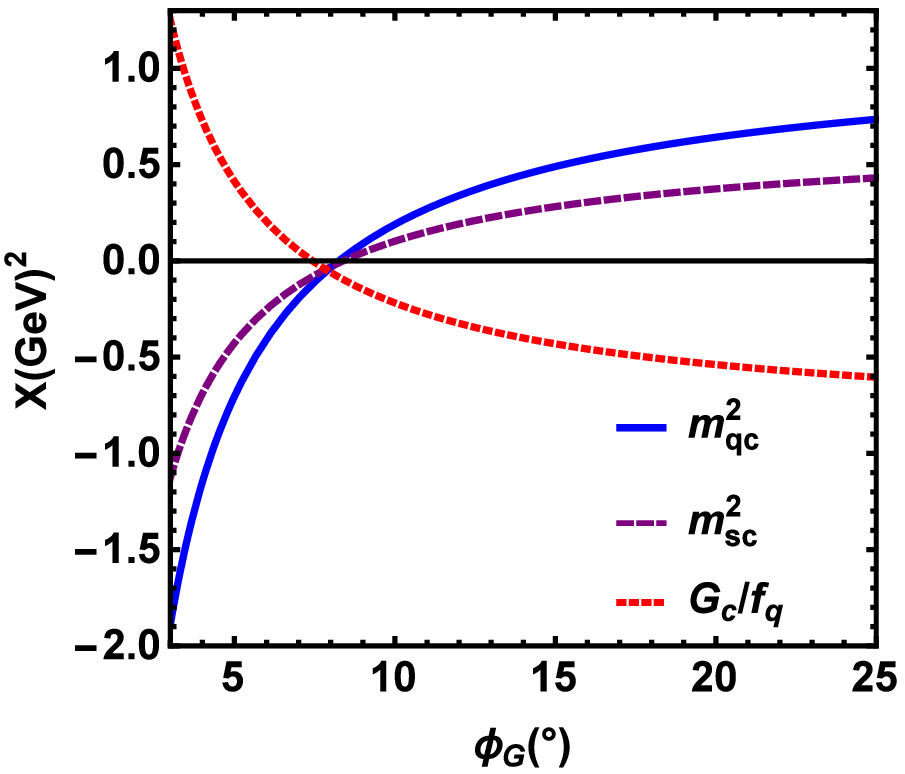}}\hspace{1.5cm}\subfigure{\includegraphics[width=5cm, height=4cm]{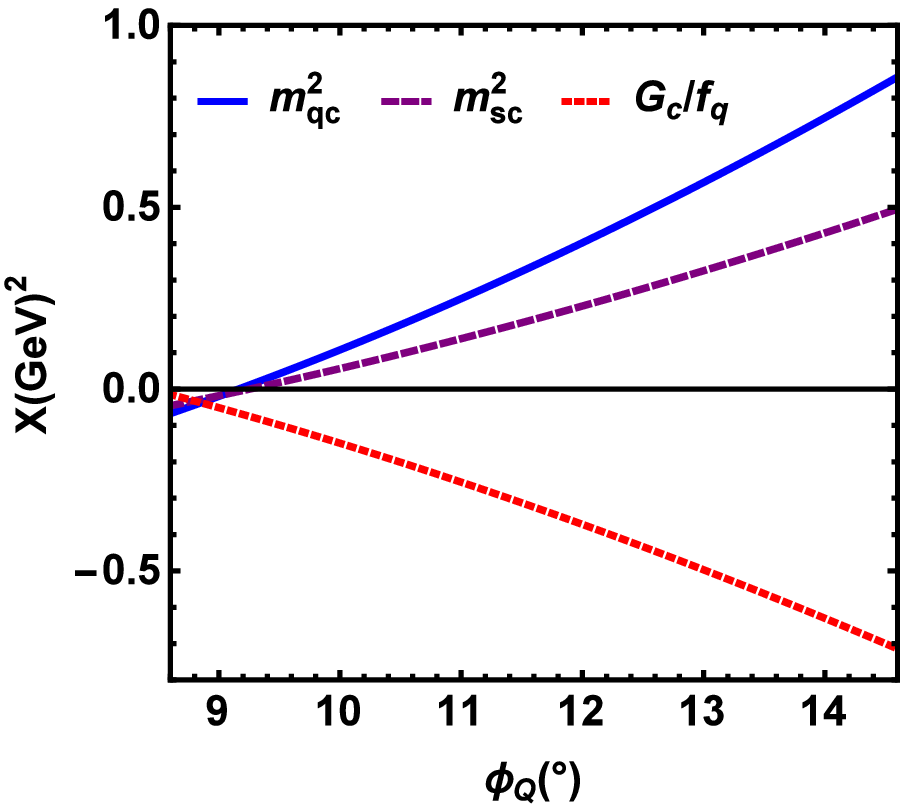}}
 \vspace{0cm}
\caption{The dependence of $m_{qc}^2$, $m_{sc}^2$ and $G_c/f_q$ on $\phi_G$ and $\phi_Q$. The results on the left panel are extracted with $m_{cc}^2=(M_{\eta_c}^2-100 \mev^2)$ and $\phi_Q=11.6^\circ$, while the results on the right are with $\phi_G=12^\circ,m_{cc}^2=(M_{\eta_c}^2-100 \mev^2)$.}\label{mqc-msc-88}
\end{figure}

\begin{figure}[htb]
\centering
\includegraphics[width=10cm, height=11cm]{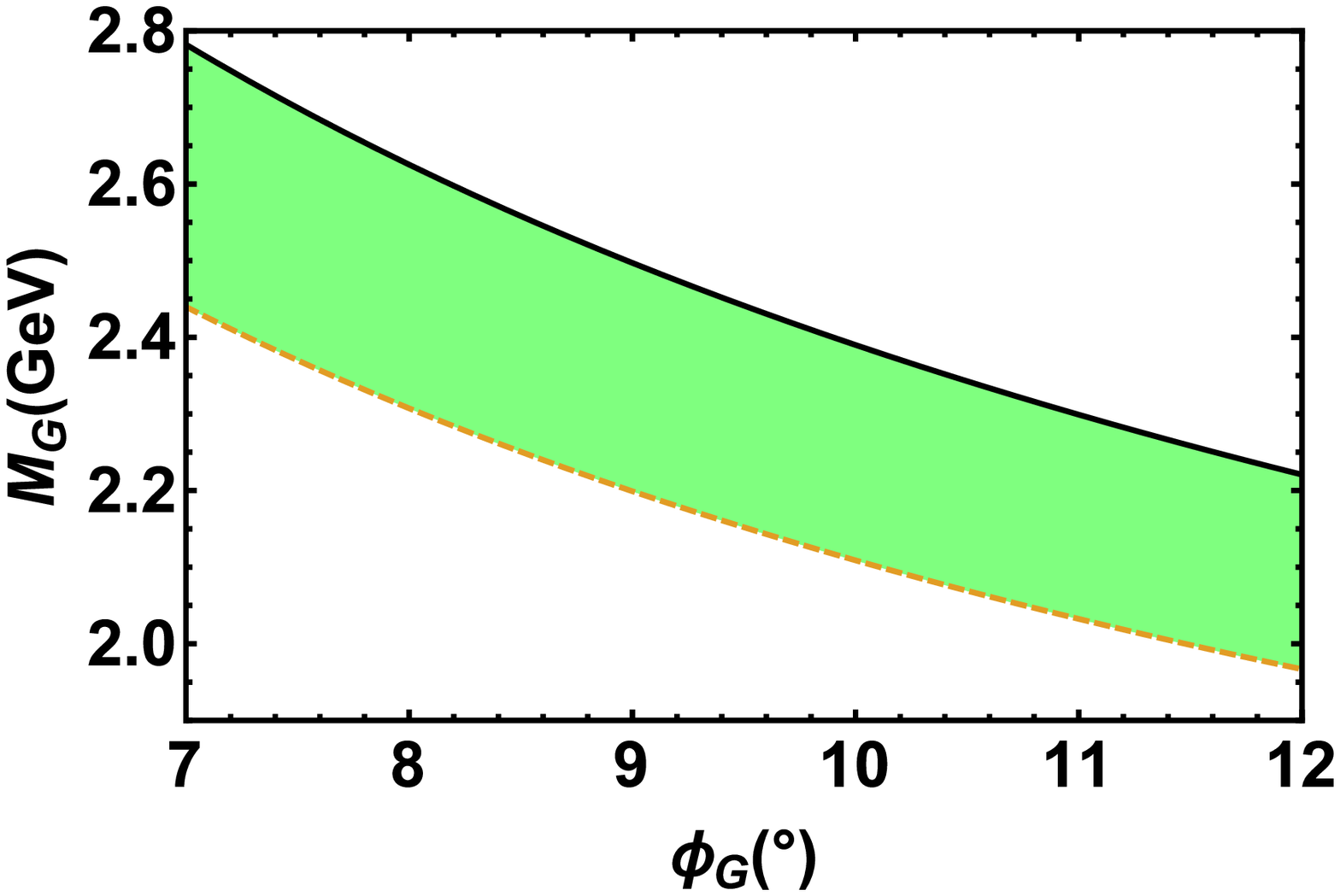}
 \vspace{-2cm}
\caption{The behavior of $M_G$ within the range of $\phi_G \in (7,12)^\circ$.}\label{MG712}
\end{figure}

\subsection{Constraints on the charmonium state}

In Table~\ref{parameters-value} $m_{qq}^2$ and $m_{ss}^2$ are of the typical values as those extracted in Refs.~\cite{Cheng:2008ss,Tsai:2011dp}. But other mass terms are rather different. As emphasized earlier, our strategy of fixing these parameters is to retain the mass hierarchy $|m_{cc}^2|\gg |m_{cg}^2|\gg |m_{cq,cs}^2|$, and $m_{ss}^2\gg m_{sg}^2$. However, $m_{qc}^2$ and $m_{sc}^2$, labelled with ``$*$" in Table~\ref{parameters-value}, seem to be abnormally large. Since $m_{qc}^2$ and $m_{sc}^2$ are determined by $\mathcal{M}_{qsgc}^{41}$ and $\mathcal{M}_{qsgc}^{42}$, it is necessary to examine the influence of the approximation that we have implemented in the determination of the unitary transformation matrix $U$ as defined in Eq.~(\ref{Umatrix}). By taking $\phi_G=12^\circ$ as an input, the $U$ matrix can be written as,
\begin{equation}\label{Umatrixvalue}
U=\left(
\begin{array}{cccc}
 0.720 & -0.693 & 0.039 &
   0.008 \\
 0.673 & 0.710 & 0.200 &
   0.041 \\
 -0.170 & -0.120 & 0.960 &
   0.197 \\
 0 & 0 & -0.201 & 0.980 \\
\end{array}
\right) \ ,
\end{equation}
where elements $U_{41}$ and $U_{42}$ are directly dropped as they are treated as small quantities. We now bring back these two elements by defining $U_{41}=x$ and $U_{42}=y$, and investigate the effects on $\mathcal{M}_{qsgc}^{41}$ and $\mathcal{M}_{qsgc}^{42}$ due to the nonvanishing $x$ and $y$. The mass matrix $\mathcal{M}_{qsgc}$ with $x$ and $y$ as explicit parameters has the following expression,
\begin{equation}
\left(
\begin{array}{cccc}
 0.699 & 0.379 & -0.583 &
   -0.120 \\
 0.379 & 0.671 & -0.383 &
   -0.0787 \\
 -1.648 x+0.076 y-0.583 &
   0.076 x-1.654 y-0.383 &
   -0.117 x-0.077 y+4.434 &
   -0.024 x-0.016 y-0.916 \\
 8.035 x-0.371 y-0.120 & -0.371
   x+8.062 y-0.079 & 0.571
   x+0.376 y-0.916 & 0.117
   x+0.077 y+8.713 \\
\end{array}
\right) \ .
\end{equation}
If we assumed that $x, \ y$ have the same order of magnitude but an opposite sign to the corresponding symmetry matrix elements in the $U$ matrix in Eq.~(\ref{Umatrixvalue}), then both $x, \ y$ should be order of $-0.01$. This is truly very small compared to the other elements. Meanwhile, the two elements $\mathcal{M}_{qsgc}^{41}$ and $\mathcal{M}_{qsgc}^{42}$ can be expressed as,
\begin{eqnarray}\label{xy}
\mathcal{M}_{qsgc}^{41} &=& m_{qc}^2+\sqrt 2 G_c/f_q \nonumber\\
                          &=&8.035 x-0.371 y-0.120  \ ,
\end{eqnarray}
and
\begin{eqnarray}
\mathcal{M}_{qsgc}^{42} &=& m_{sc}^2+ G_c/f_s\nonumber\\
                          &=&-0.371x+8.062 y-0.079 \ .
\end{eqnarray}
The above two equations suggest that to keep both $\mathcal{M}_{qsgc}^{41}$ and $\mathcal{M}_{qsgc}^{42}$ small it requires intrinsic dynamic constraints on both $m_{qc}^2$ and $m_{sc}^2$ of which the effects will then show up via $x$ and $y$ in the $U$ matrix. The dependence of $m_{qc}^2$ and $m_{sc}^2$ on $m_{cc}^2$, $\phi_G$ and $\phi_Q$ are encoded via Eq.~(\ref{parameters}). It means that more stringent constraints on $m_{cc}^2$, $\phi_G$ and $\phi_Q$ should be applied in order to keep both $m_{qc}^2$ and $m_{sc}^2$ small. Interestingly, by requiring $|m_{qc},m_{sc}|\leq 0.25$ GeV$^2$, we find that $m_{cc}^2$ should be restricted within $(8.7\sim 8.8)$ GeV$^2$ as shown by Fig.~\ref{mqc-msc}. It corresponds to $(M_{\eta_c}-m_{cc})=13.5\sim 30.4$ MeV which means that the glueball-$c\bar{c}$ mixing has resulted in a mass gap between the physical and pure states. Meanwhile, it shows that the glueball-$c\bar{c}$ mixing does not change the main character of $\eta_c$ as the ground state pseudoscalar charmonium. However, due to the mixing, some of the observables may have indicated effects arising from a small glueball component in the wavefunction of $\eta_c$. This is consistent with the conclusion of Ref.~\cite{Tsai:2011dp}.

To show the sensitivities of  $m_{qc}^2$, $m_{sc}^2$ and $G_c/f_q$ to $m_{cc}$, we set the value of $m_{cc}^2$ to be $100 \ \mev^2$ below the mass of $\eta_c$ squared, i.e. $m_{cc}^2=M_{\eta_c}^2-100 \ \mev^2$. As shown by Fig.~\ref{mqc-msc-88}, $\phi_G$ will be restricted within $(7,12)^\circ$ and $\phi_Q$ cannot be larger than $11.6^\circ$.

\subsection{Pseudoscalar glueball production in $J/\psi$ radiative decay}

In Fig.~\ref{MG712} the glueball mass in term of the favored range for $\phi_G$ is plotted. The band indicates the boundary of $G_g$ with $(-0.054 \pm 0.008)$ GeV$^3$.  It should be noted that, if $M_G=2.56 \ \gev$ from the LQCD calculation~\cite{Sun:2017ipk} is taken, $\phi_G$ would be fixed as $7^\circ$, and the corresponding $U$ matrix is given as,
\begin{equation}
U=\left(
\begin{array}{cccc}
 0.72 & -0.69 & 0.023 & 0.0047 \\
 0.68 & 0.72 & 0.12 & 0.024 \\
 -0.099 & -0.070 & 0.97 & 0.20 \\
0 & 0 & -0.20 & 0.98 \\
\end{array}
\right) \ .
\end{equation}

The production rate of $P$ in $J/\psi \to \gamma P$ scales as $(G_P/\bar{M}^2)^2$~\cite{Cheng:2008ss}, where $\bar{M}$ is a typical energy scale for $\langle 0|\alpha_s(\bar{M}) G{\tilde G}/(4\pi)|(q\bar{q})_{0^{-+}}\rangle$. This energy scale also determines the strong coupling $\alpha_s(\bar{M})$. It is natural to expect that this energy scale is the same for the light pseudoscalar meson productions, i.e. $\eta$ and $\eta'$, in the light flavor sector. However, it should be different for $\eta_c$ due to the much shorter range for the color force between $c$ and $\bar{c}$ and larger momentum transfers to the gluons in the $c\bar{c}\to gg$ transition. We can examine the fitted values in Eq.~(\ref{suscept}) for $\eta$ and $\eta'$, and then estimate the production rate for the pseudoscalar glueball.

The branching ratio fraction for the production of two pseudoscalar mesons $P_1$ and $P_2$ in the $J/\psi$ radiative decays can be expressed as
\begin{equation}
\frac{BR(J/\psi\to\gamma P_1)}{BR(J/\psi\to\gamma P_2)}=\left(\frac{G_{P_1}}{G_{P_2}}\right)^2\left(\frac{\bar{M}_2}{\bar{M}_1}\right)^4\left(\frac{q_1}{q_2}\right)^3 \ ,
\end{equation}
where $q_1$ and $q_2$ are the three-vector momentum of the pseudoscalar meson $P_1$ and $P_2$ in the rest frame of $J/\psi$, respectively, while $\bar{M}_1$ and $\bar{M}_2$ are the energy scales for the strong quark-gluon couplings for $P_1$ and $P_2$, respectively, in the $q\bar{q}\to gg$ transition. As mentioned earlier, it is a reasonable approximation to adopt the same $\bar{M}$ value for $\eta$ and $\eta'$.

With the data for ${BR(J/\psi\to\gamma \eta')}$ and ${BR(J/\psi\to\gamma \eta)}$ from experiment~\cite{Patrignani:2016xqp} it allows us to extract $G_{\eta'}/G_\eta=2.39^{+0.08}_{-0.15}$, where the central values of the data give the ratio, and the boundaries are given by the upper and lower limit of the data uncertainties~\cite{Patrignani:2016xqp}. The theoretical value from Eq.~(\ref{suscept}) gives $G_{\eta'}/G_\eta=3.19$ which is close to the data constraint taking into account the uncertainties arising from the parameters. The branching ratio fraction can be related to the $\eta$-$\eta'$ mixing either with or without the glueball mixing which indicates the small glueball component in the $\eta$ and $\eta'$ wavefunction as found in the literature~\cite{Feldmann:1998vh,Feldmann:1998sh,Feldmann:1999uf,Li:2007ky,Escribano:2007cd}.

For the glueball production in $J/\psi$ radiative decays, one can calibrate its production to the rate for $J/\psi\to \gamma\eta$ via
\begin{equation}
\frac{BR(J/\psi\to\gamma G)}{q_G^3}=\left(\frac{G_G/\bar{M}_g^2}{G_\eta/\bar{M}_\eta^2}\right)^2\frac{BR(J/\psi\to\gamma\eta)}{q_\eta^3} \ ,
\end{equation}
where $\bar{M}_g$ and $\bar{M}_\eta$ are the energy scales for the glueball and $\eta$. Note that in case there is a large mass difference between the physical glueball mass and $\eta$ it is unnecessary for $\bar{M}_g=\bar{M}_\eta$. Early studies of the scale relation can be found in Ref.~\cite{Kataev:1981aw}. As an approximation we assume $\bar{M}_g=\bar{M}_\eta$ to extract $BR(J/\psi\to\gamma G)$ with a mass of $M_G=2.1$ GeV, i.e. $BR(J/\psi\to\gamma G)\simeq 3.8\times 10^{-3}$. Taking into account that $\bar{M}_g$ is supposed to be larger than $\bar{M}_\eta$, this rate sets up an upper limit to the branching ratio for the production of an $M_G=2.1$ GeV pseudoscalar glueball in the $J/\psi$ radiative decays. This result is consistent with the analysis of Ref.~\cite{Gabadadze:1997zc} which pointed out the difficulty of reconciling the LQCD result with the experimental hint for the possible existence of the additional $\eta(1405)$.

One notices that in this mass region BESII reported pseudoscalar states $X(2120)$ and $X(2370)$ in the invariant mass spectrum of $\eta'\pi\pi$ in $J/\psi\to \gamma X\to \gamma \eta'\pi\pi$~\cite{Ablikim:2005um,Liu:2010tr}, which was confirmed by BESIII later with high statistics~\cite{Ablikim:2016itz}. The PDG also list $\eta(2225)$ as an established state in $J/\psi\to\gamma K\bar{K}\pi$ with $BR(J/\psi\to \gamma\eta(2225)=(3.14^{+0.50}_{-0.19})\times 10^{-4}$~\cite{Patrignani:2016xqp}. Whether these states are radial excitations of $\eta$ and $\eta'$~\cite{Yu:2011ta} or whether one of these states is the pseudoscalar glueball candidate should be further investigated in both experiment and theory.

\section{Summary}\label{summary}

In this work, we revisit the mechanism proposed and studied in Refs.~\cite{Tsai:2011dp,Cheng:2008ss} for the pseudoscalar meson and glueball mixings. On the one hand, we confirm many results from Refs.~\cite{Tsai:2011dp,Cheng:2008ss} on the correlations among the introduced parameters. On the other hand, we scrutinize the dynamical constraints on the glueball mass and clarify that the physically favored parameter space would lead to much higher glueball mass than that obtained before. In particular, we show that the approximation of neglecting both $m_{qg}^2$ and $m_{sg}^2$ in the extraction of the glueball mass was inappropriate. Although $m_{qg}^2$ is indeed a negligible quantity in comparison with $\sqrt{2}G_g/f_q$, the value of $m_{sq}^2$ is actually comparable with $G_g/f_s$ and cannot be neglected. After properly treat these parameters and identify $G_g$ and $\phi_G$ as the parameters that play a dominant role in the determination of the mixing pattern, we find that the glueball mass $M_G$ cannot be lower than 1.9 GeV which is much higher than 1.4 GeV determined by the approximation in Refs.~\cite{Tsai:2011dp,Cheng:2008ss}. We find that the mixing angle $\phi_G \in (7,12)^\circ$ for the glueball and light flavor states is favored in our model. It allows the estimate of an upper limit branching ratio of the production of pseudoscalar glueball in $J/\psi$ radiative decay.

This result is encouraging in such a sense that it resolves not only the apparent paradox between the LQCD results and some old experimental data for the pseudoscalar glueball mass, but also explains the single peak structure around 1.4$\sim$1.5 GeV observed by the recent high-statistics measurements at BESIII in exclusive decay channels~\cite{Ablikim:2016hlu,Ablikim:2010au,Liu:2010tr}, although in some channels the peak positions are slightly shifted. The peak position shift is due to interferences from the triangle singularity mechanism~\cite{Wu:2011yx,Wu:2012pg,Liu:2015taa} instead of by two-pole structures. Namely, we can conclude that there is no need for two light pseudoscalars $\eta(1405)$ and $\eta(1475)$ to be present as two individual states and with the $\eta(1405)$ as the pseudoscalar glueball candidate.  Consequently, as pointed out in Refs.~\cite{Wu:2011yx,Wu:2012pg,Liu:2015taa}, in order to search for the pseudoscalar glueball candidate one should look at the higher mass region at least above 1.8 GeV where some of the recently observed pseudoscalar states by BESIII~\cite{Ablikim:2016hlu} should be carefully examined.

\section{Acknowledgement}

The authors thank Ying Chen, Hsiang-nan Li, and K.-F. Liu for useful discussions. This work is supported, in part, by the National Natural Science Foundation of China (NSFC) under Grant Nos. 11425525, 11521505, 11375061, and 11775078, by DFG and NSFC through funds provided to the Sino-German CRC 110 ``Symmetries and the Emergence of Structure in QCD'' (NSFC Grant No. 11261130311), and by the National Key Basic Research Program of China under Contract No.~2015CB856700.


\end{document}